\documentclass[prd,twocolumn,showpacs,preprintnumbers,amsmath,amssymb, floatfix]{revtex4}

\def\msun{$M_{\odot}$}
\usepackage{graphicx}
\usepackage{dcolumn}
\usepackage{bm}

\begin{document}

\newcommand{\btau}{\mbox{\boldmath{$\tau$}}}

\newcommand{\bphi}{\mbox{\boldmath{$\phi$}}}
\newcommand{\btheta}{\mbox{\boldmath{$\theta$}}}
\newcommand{\bphisc}{\mbox{\small\boldmath{$\phi$}}}

\newcommand{\comment}[1]{}

\title{Geometrical Expression for the Angular Resolution of a Network of Gravitational-Wave Detectors}

\author{Linqing  Wen}
\email{lwen@cyllene.uwa.edu.au}
\affiliation{International Center for Radio Astronomy Research,  School of Physics, University of Western Australia, 35 Stirling Hwy, Crawley, WA 6009, Australia}
\author{Yanbei Chen}
\email{yanbei@tapir.caltech.edu}
 \affiliation{
Division of Physics, Mathematics, and Astronomy, Caltech, Pasadena, CA 91125, USA}



\begin{abstract}  
We report for the first time general geometrical expressions for the angular resolution of an arbitrary network of interferometric gravitational-wave (GW) detectors when the arrival-time of a GW is unknown.  We show explicitly elements that decide the angular resolution of a GW detector network.   In particular, we show the dependence of the angular resolution on areas formed by projections of pairs of detectors and how they are weighted by sensitivities of individual detectors.  Numerical simulations are used to demonstrate the capabilities of the current GW detector network.   We confirm that the angular resolution is poor along the plane formed by current LIGO-Virgo detectors.  A factor of a few to more than ten fold improvement of the angular resolution can be achieved if the proposed new GW detectors LCGT or AIGO are added to the network.  We also discuss the implications of our results for the design of a GW detector network,  optimal localization methods for a given network, and electromagnetic follow-up observations.
\end{abstract}
\pacs{04.80.Nn, 95.75.-z, 97.80.-d, 97.60.Gb} 

\maketitle

\section{Introduction}
\label{intro}

Several types of astrophysical sources are expected to be detectable both in gravitational waves (GWs) and in conventional electromagnetic (EM) wavelengths. 
For example, long gamma-ray bursts have been conjectured to originate from asymmetric core collapse of massive stars, and short gamma-ray bursts might be produced by the coalescence of compact binary objects containing neutron stars.  Both of these could emit gravitational waves in the frequency band of ground-based laser interferometer GW detectors (e.g., Ref.~\cite{grb_gw}).   Several large-scale interferometric GW detectors have reached (or approached) their design sensitivity, and are coordinating to operate as a global array.  These include the LIGO detectors at Livingston, Louisiana, and Hanford, Washington, US, the Virgo detector in Pisa, Italy, the GEO\,600 detector in Hannover, Germany, and the TAMA\,300 detector in Tokyo Japan.  Upgrades to existing detectors (Advanced LIGO and Advanced Virgo) have been planned \cite{ligo,virgo,geo}, while new detectors (LCGT in Japan \cite{lcgt} and AIGO in Australia \cite{aigo}) are still being proposed.   In case of a strong EM event, follow-up searches for GW signals can be conducted in archived data in the time window of the event (e.g., Ref.~\cite{ligo_grb}). On the other hand, EM follow-ups to probable GW events require a clear understanding of the angular resolution of an array of GW detectors.   

The angular resolution of an individual GW detector, arising from its antenna beam pattern,  is
rather poor \citep{thorne87}.  However, the large baselines of the current
GW-detector network facilitate better angular resolution via
triangulation.  Several localization algorithms have been proposed and the effect of arrival timing uncertainties as well as amplitude information of GWs have been investigated \cite{tinto89, wen05a, cavalier06, beauville06, acernese07, wen07a, wen_fan, markowitz08,searle08,searle09, steve09}.  Quantitative studies of the angular resolution of a network of GW detectors have been conducted by
several authors, both for a ground-based detector network and for the
future space GW detector LISA \citep{krolak94,cutler98,Pai01,leor04}.
A standard approach is to calculate numerically the Fisher information
matrix, which leads to a method-independent lower bound on the statistical
errors of estimated parameters (see a review in Ref.~\cite{krolak05}).
On the other hand, explicit analytical expressions for the network angular
resolution are rare in the literature largely because of the
complexity involved in derivations.  

Two approximate analytical expressions for the angular resolution can be
found in the literature (summarized in \citep{sylvestre03}) for a network of three GW detectors.  One is an elegant approximate geometrical formula for 3 detectors due to Thorne (as cited in Eq.(8.3) of Ref.~\cite{tinto89}):
the solid angle uncertainty is
\begin{equation}
\Delta \Omega = \frac{2c^2 \Delta \tau_{12}\Delta \tau_{13}}{A \cos \theta}\,,
\end{equation}
where $c$ is the speed of light, $\Delta \tau_{12}$ and $\Delta \tau_{13}$, are time-of-arrival accuracy between pairs of detectors,  $A$ is the area formed by the three detectors, and $\theta$ is the angle between the source direction and the normal to the plane of the three detectors.  However, the underlying assumptions and derivation of this expression are not available in the literature.   The dependence of angular resolution on the signal-to-noise ratio (SNR) was derived by Tinto in Ref.~\citep{tinto89}, by expressing the above time-of-arrival accuracy as a function of SNR and frequency~\citep{schutz89} (derived from the Fisher matrix assuming all other
information of the waveform as perfectly known).  The other formula is based on the
numerical result of the angular resolution of the LIGO-Virgo network around a wave incident direction normal to the plane formed by the three detector sites~\citep{sylvestre03} ---  for GWs emitted from neutron star-neutron star (NS-NS) inspirals~\citep{Pai01}.   This particular resolution was then rescaled by the cosine
of the wave incidence angle and SNR~\citep{sylvestre03}.  Analytical geometrical expressions or approximate ones for the angular
resolution for an array of more than three detectors have not been obtained in the literature. 

In this paper, we deduce explicit analytical expressions for the angular resolution of an arbitrary GW detector network in terms of observables such as cross-sectional areas of the network and energy flux of the incoming GW.   We use only the time-of-arrival information, ignoring additional (usually rather poor) information from the directional-derivatives of antenna beam pattern functions --- and therefore arrive at a conservative estimate.  Such an approximation allows us to obtain expressions that have explicit geometrical meanings, further generalizing Thorne's formula to an arbitrary number of detectors, and several particular scenarios.  In particular, we consider both {\it short signals,} during which motion of the detector network is negligible, and {\it long signals,}  for which the trajectory traced by the detectors during the signal determines the effective size of the detector network.  We also consider signals with  {\it  known} or {\it unknown} waveforms,  but always assume unknown arrival times of the signal.  In this paper, the scenarios where signals have completely {\it known} and completely {\it unknown} waveforms are termed interchangeably as the {\it best-case } and the {\it worst-case} scenarios respectively.      

We focus on deriving explicit expressions for several situations that will arise in the practice of searching for and localizing GWs.   Specifically, we derive general expressions for {\it short signals} assuming only that arrival time is one of the unknown parameters (summarized in Eq.~\eqref{Omega} and text thereafter)  and for {\it long signals} assuming known waveform (Eq.~\eqref{long_general}).  Based on these formulae,  we show simplified solutions for Eq.~\eqref{Omega} in several realistic situations: (1) {\it short signals} in the {\it worst-case} scenario (Eq.~\eqref{worst}) and in the {\it best-case} scenario (Eq.~\eqref{best}) for an arbitrary network of detectors,  (2) special cases of the two- and three- detector networks in the {\it best case} (Eq.~\eqref{theta_2}--\eqref{ang_3det}) and their representations when the wave is short and monochromatic (Eq.~\eqref{theta_2f}--\eqref{omega_3}), (3) {\it long signals} in short and long observations with detectors in circular motion (Eq.~\eqref{pulsar_short} and Eq.~\eqref{pulsar_long} respectively).  



This paper is organized as follows.  In section~\ref{principle}, we explain our notation.  In section~\ref{expression},  we derive analytical expressions of the angular resolution for an arbitrary detector network and for special cases.   We show explicit derivations for the {\it worst-case} and the {\it best-case} scenario in section~\ref{sec_worst} and section ~\ref{sec_best} and then derive a general expression in section~\ref{short_general}.   In section~\ref{sec_long}, we derive a general expression for long-duration wave and its application to detectors at circular motion.  We then discuss the implications for the design of a detector network and localization strategies in section~\ref{implication}.  The astrophysical applications of our results are shown in section~\ref{astro}.  In section~\ref{sec_antenna}, we discuss the possible errors in our estimation caused by ignoring the directional derivatives of antenna beam patterns.   Our results are summarized in section~\ref{concl}. 

\section{Mathematical Preliminaries}
\label{principle}

\subsection{Antenna Pattern of a Detector Network}

In this paper, we assume a network of $N_d$ gravitational-wave detectors, with spatial locations given by the vector $\mathbf{r}_I$ with $I=1,2,\ldots N_d$, each with spatial size much smaller than the GW wavelength.  The strain of an incoming GW observed by an individual detector $I$ is then a linear combination of the two wave  polarizations in the transverse traceless gauge,
\begin{equation}
d_I(t_0+\tau_I+t)=f^{+}_Ih_{+}(t)+f^{\times}_Ih_{\times}(t),\quad 0<t<T\;,
\label{d_t}
\end{equation}
where  $t_0$ is the arrival time of the wave at the coordinate origin and $\tau_{I}$ is the time required for the wave to travel from the origin to reach the $I$-th  detector at time $t$, 
\begin{equation}
\tau_I (t) = {\mathbf n} \cdot \mathbf{r}_I(t)/c\; . 
\label{tau_I}
\end{equation}
Here  ${\mathbf{n}}$ is the propagation direction of a GW, $t  \in [0,T]$ is the time label of the wave, and $T$ is the signal duration.  The quantities  $f^{+}$ and $f^{\times}$ are the detector's {\it antenna beam pattern} functions \citep{krolak98} for the two wave polarizations ($h_+$, $h_{\times}$).  They depend on the relative orientation between the detector configuration and the frame in which the polarizations are defined (which is in turn related to the source direction ${-\mathbf{n}}$).   In particular, given a Michelson-type interferometer with orthogonal arms along $\mathbf{e}_x$ and $\mathbf{e}_y$, and given the symmetric, trace-free polarization tensors $\mathbf{e}_+({\mathbf{n}})$ and $\mathbf{e}_\times({\mathbf{n}})$ to which the wave polarizations refer, we have
\begin{equation}
f_{+,\times}({\mathbf{n}}) =
(\mathbf{e}_x\otimes\mathbf{e}_x- \mathbf{e}_y\otimes\mathbf{e}_y):\mathbf{e}_{+,\times}({\mathbf{n}})
\end{equation}
where the symbol $:$ stands for contraction.  Note that different conventions can be used to define $\mathbf{e}_{+,\times}({\mathbf{n}})$, as long as these tensors are symmetric,  trace-free, and satisfy 
\begin{equation}
{\mathbf{n}} \cdot\mathbf{e}_{+,\times}({\mathbf{n}})=0
\end{equation}

The Fourier transform of the time-series data from the $I$-th GW detector is
\begin{equation}
d_I(\Omega) = \int^T_0 d_I(t) e^{i\Omega t} dt\; .  
\end{equation}
Denoting the corresponding one-sided noise spectral density by $S_I (\Omega)$,  we define a whitened data set in the frequency domain,
\begin{equation}
\hat{{d_I}} (\Omega) = S_I^{-\frac{1}{2}} (\Omega) d_I(\Omega)\,.\quad 
\end{equation}
Vector $\hat{\mathbf{d}}(\Omega)$ then corresponds to the whitened data set at each frequency.  For short-duration signals where motion of the detector array is unimportant, antenna beam patterns are treated as constant, hence 
\begin{equation}
\label{dwhite}
\hat{\mathbf{d}}(\Omega)  = e^{-i\bphi}e^{-i\Omega t_0}
\hat{ \mathbf{A}} \mathbf{h}(\Omega)\,,
\end{equation}
where $\bphi$ is a $N_d\times N_d$ diagonal matrix  with 
$\phi_{IJ} \equiv  \delta_{IJ} \Omega\tau_J$, or
\begin{equation}
\bphi =\Omega  \left[ \begin{array}{ccc}
 \tau_1 \\
& \ddots & \\
& &\tau_{N_d}
\end{array}
\right]
=\Omega \left[ \begin{array}{ccc}
 \frac{{\mathbf{n}}\cdot\mathbf{r}_1}{c} \\
& \ddots & \\
& &  \frac{{\mathbf{n}}\cdot\mathbf{r}_{N_d}}{c}
\end{array}
\right],
\end{equation}
$\hat{\mathbf{A}}$ is an $N_d \times 2$ matrix of the antenna pattern functions for all detectors weighted by noise,
\begin{equation}
\hat{\mathbf {A}} \equiv 
\left[
\begin{array}{cc}
\frac{f^+_1({\mathbf{n}})}{\sqrt{S_1(\Omega)}} & \frac{f^\times_1({\mathbf{n}})}{{\sqrt{S_1(\Omega)}}} \\
\vdots & \vdots \\
\frac{f^+_{N_d}({\mathbf{n}})}{\sqrt{S_{N_d}(\Omega)}} & \frac{f^\times_{N_d}({\mathbf{n}})}{\sqrt{S_{N_d}(\Omega)}}
\end{array}
\right],
\label{A_hat}
\end{equation}
and $\mathbf{h}(\Omega)$ is a 2-dimensional vector function containing the two polarizations of a GW in the frequency domain,
\begin{equation}
\mathbf{h}(\Omega)
=
\left[
\begin{array}{c}
h_+(\Omega) \\
h_\times(\Omega)
\end{array}
\right]\,.
\end{equation}

For simplicity, we keep the $\Omega$-dependence in the notation only when
it is necessary for clarity.  At any individual frequency $\Omega$, even though the response vector $\hat{\mathbf{d}}$ is $N_d$ dimensional, the existence of only two independent signal polarizations, $+$ and $\times$, means that the set of all possible signal vectors is 2-dimensional. 

\subsection{Fisher Matrix}
We define the following inner product between two vectors, 
\begin{eqnarray}
\langle \mathbf{a} | \mathbf{b} \rangle  &=& 2 \int_{-\infty}^{+\infty}\frac{d\Omega}{2\pi}{\mathbf{a^\dagger}
\mathbf{b}} \nonumber \\
&=& 2 \sum_{I}
\int_{-\infty}^{+\infty}\frac{d\Omega}{2\pi}{a^{*}_I(\Omega) 
b_I(\Omega)}. \label{braket2} \end{eqnarray}
Under the assumption of stationary Gaussian detector noise, the optimal squared signal-to-noise ratio is \cite{finn92},
\begin{eqnarray}
\rho^2_{N}=\langle \hat {\mathbf{{d}}} | \hat{\mathbf{{d}}} \rangle.
\label{rho_SNR}
\end{eqnarray}

The Fisher matrix for parameters $\vec\theta$ can be defined as
\begin{equation}
\Gamma_{ij} =\langle \partial_{\theta_i} \hat{\mathbf{d}}| \partial_{\theta_j} \hat{\mathbf{d}}\rangle\; ;
\end{equation}
note that data are already whitened.  The Cramer-Rao bound \cite{cramer46} states that for an unbiased estimator (the ensemble average of which is the true value), the Fisher matrix sets a method-independent lower bound for the covariance matrix of estimated parameters when considering statistical errors.  In the case of high signal-to-noise ratio,  the covariance matrix of a set of parameters $\vec \theta$ is approximately given by the inverse of the Fisher matrix (see \cite{cutler98, leor04, michele08, tom08} and references therein), 
\begin{equation}
V_{ij} =({\mathbf{\Gamma}}^{-1})_{ij}.
\end{equation}
Now suppose that in addition to $\vec\theta$, there exist more unknown parameters, $\vec\lambda$.  Then the Fisher matrix pertaining to $\vec\theta$ can be written as
\begin{equation}
\min_{\delta \vec\lambda} \langle \delta \hat{\mathbf{d}}  | \delta\hat{ \mathbf{d}} \rangle = \frac{1}{2}\Gamma_{ij}\delta\theta_i \delta\theta_j + O(|\delta\vec\theta|^3)\,,
\end{equation}
where
\begin{eqnarray}
\delta \hat{\mathbf{d}} \equiv  \hat{\mathbf{d}}(\vec\theta+\delta\vec\theta,\vec\lambda+\delta \vec\lambda) -\hat{\mathbf{d}}(\vec\theta,\vec\lambda)\; .
 \end{eqnarray}
Using this formulation, $\vec\lambda$ can also consist of a continuum of parameters --- the minimization would become a quadratic variational problem.  Note that for a discrete parameter set, the covariance matrix can also be calculated using standard matrix inversion for a matrix consisting of block matrices.  In this paper, we show our derivations based on the variational method. 

\section{Analytical Expressions of Angular Resolution}
\label{expression}
We calculate the angular resolution of a detector network by applying the Fisher matrix to obtain method-independent lower limits \cite{cramer46} on the statistical errors in estimating the direction of a GW source.  The limits are for unbiased estimators and for Gaussian noise (for cautions in using these limits, see \cite{michele08}).  

We denote $Z$ as the direction of $\mathbf{n}$, and then the error in solid angle (measured in steradians) is defined by covariance of $n_X$ and $n_Y$ as
\begin{equation}
\Delta\Omega_s = 2 \pi \sqrt{\langle\delta n_X^2\rangle \langle \delta n_Y^2\rangle - \langle\delta n_X \delta n_Y\rangle^2} =2 \pi (\det \mbox{\boldmath$\Gamma$})^{-1/2}\,,
\label{Omega_def}
\end{equation}
where $\delta n_X$ and $\delta n_Y$ are the deviations of $n_X$ and $n_Y$ from their true values and {\boldmath$\Gamma$} is the Fisher matrix pertaining to angular parameters $n_X$ and $n_Y$.  The factor of $2\pi$ is introduced so that the probability that estimated parameters fall outside an area $\Delta \Omega$ due to statistical error is $e^{- \Delta \Omega/\Delta \Omega_s}$.  For angular parameters in a polar coordinate system with colatitude $\theta$ and longitude $\phi$, we have   
$\Delta\Omega_s = 2\pi |\cos{\theta}| \sqrt{\langle \Delta \theta^2\rangle \langle\Delta \phi^2\rangle - \langle\Delta \theta \Delta \phi\rangle^2} $. 
In the next two subsections, we derive expressions for $\det \mbox{\boldmath $\Gamma$}$ and $\Delta\Omega_s$ for {\it short} and {\it long} signals.

\subsection{Short-duration GWs}
\subsubsection{Worst-case scenario: signal with unknown waveform}
\label{sec_worst}
For a {\it short signal} with unknown waveform, the unknown parameters consist of the sky coordinates of the gravitational-wave source $n_j$ ($j=1,2$) and the unknown waveform $\mathbf{h}(\Omega)$.  Varying all unknown parameters, we have (note that $\delta t_0$ is absorbed into $\delta \mathbf{h}$), 
\begin{equation}
\delta \hat{\mathbf{d}} (\Omega) = e^{-i\bphi}\left[\left(-i \sum_j{\partial_j \bphi} \hat{\mathbf{A}} \mathbf{h}(\Omega) \delta n_j\right)   +\hat{\mathbf{A}}\delta \mathbf{h}(\Omega)\right],
\label{dd_dn}
\end{equation}
where  we have ignored the change of $\hat{\mathbf{A}}$ induced by $\delta {n_j}$ (see sec.~\ref{sec_antenna} for a discussion of its effect) and defined
\begin{equation}
( \partial_j\bphi )\equiv\frac{\partial\bphi}{\partial n_j}
\end{equation}
In the following, recall that $\hat{\mathbf{d}}$, $\delta\hat{\mathbf{d}}$, $\mathbf{h}$, $\delta\mathbf{h}$, $\bphi$ and $\hat{\mathbf{A}}$ are frequency dependent, but we drop their frequency dependence in equations for simplicity.  We then have 
\begin{eqnarray}
\langle \delta \hat{\mathbf{d}} |  \delta\hat{\mathbf{d}}\rangle 
&= & \sum_{j,k} 
\langle (\partial_j\bphi) 
\hat{\mathbf{A}} \mathbf{h}   
 |  (\partial_k\bphi)
\hat{\mathbf{A}} \mathbf{h} \rangle \delta n_j \delta n_k  \nonumber \\
&+& i  \sum_j \langle (\partial_j \bphi)
\hat{\mathbf{A}}\mathbf{h}  
| 
\hat{\mathbf{A}}\delta \mathbf{h}\rangle \delta n_j  \nonumber \\
&-& i \sum_k \langle 
\hat{\mathbf{A}} \delta \mathbf{h}| 
(\partial_k \bphi)
\hat{\mathbf{A}} \mathbf{h}\rangle \delta n_k \nonumber \\
&+& \langle \hat{\mathbf{A}} \delta \mathbf{h}  | \hat{\mathbf{A}} \delta \mathbf{h} \rangle \,.
\end{eqnarray}
Minimizing $\langle \delta \mathbf{d} | \delta\mathbf{d}\rangle $ over $\delta\mathbf{h}$, we obtain 
\begin{equation}
\delta \mathbf{h} = i \sum_k (\hat{\mathbf{A}}^\dagger \hat{\mathbf{A}})^{-1} \hat{\mathbf{A}}^\dagger
(\partial_k \bphi )\hat{\mathbf{A}} \mathbf{h}\,\delta n_k\,, 
\end{equation}
and
\begin{eqnarray}
\Gamma_{jk} &=& 
\langle (\partial_j \bphi) 
\hat{\mathbf{A}}\mathbf{h}  
| 
(\mathbf{I} - \mathbf{P})
(\partial_k \bphi )
 \hat{\mathbf{A}} \mathbf{h}
\rangle\nonumber \\
&=&\langle (\partial_j \bphi)
\hat{\mathbf{d}} |  
\left(\mathbf{I} - \mathbf{P}
\right)
(\partial_k \bphi )
\hat{\mathbf{d}}\rangle
\end{eqnarray}
where we have defined $\mathbf{I}$ as the identity matrix and 
\begin{equation}
 \mathbf{P}\equiv \hat{\mathbf{A}}( \hat{\mathbf{A}}^\dagger \hat{ \mathbf{A}})^{-1}  \hat{\mathbf{A}}^{\dagger}\,,
\label{worst_P}
 \end{equation}
which is the projection operator into the whitened signal space, i.e., the 2-dimensional space spanned by the columns of $\hat{\mathbf{A}}$. 

After a straightforward algebraic manipulation, we obtain
\begin{eqnarray}
&&\left[\det\mbox{\boldmath$\Gamma$} \right] \nonumber \\
&=& \frac{1}{8c^4} \sum_{J,K,L,M} \Delta_{JK}\Delta_{LM}
|(\mathbf{r}_{KJ} \times \mathbf{r}_{ML})\cdot \mathbf{n}|^2
\label{worst_gen}
\end{eqnarray} 
where
\begin{eqnarray}
\Delta_{JK} &=&\langle \Omega \hat{d}_J | (\delta_{JK} - P_{JK})\Omega \hat{d}_K\rangle \nonumber \\
&=&(\delta_{JK} - P_{JK}) \left ( 2 \int_{-\infty}^{+\infty}\frac{d\Omega}{2\pi}{\Omega^2 \hat{d}^*_J \hat{d}_K}\right ) \,,
\label{worst}
\end{eqnarray}
and $\mathbf{r}_{JK} \equiv \mathbf{r}_K -\mathbf{r}_J$.  

Note that $|(\mathbf{r}_{KJ} \times \mathbf{r}_{ML})\cdot \mathbf{n}|$, is just twice the area formed by the projections of the detectors $J$, $K$, $L$ and $M$ onto the plane orthogonal to the wave propagation direction.   The quantity $\Delta_{JK}$ can be interpreted as the projection of the weighted data correlation between detectors $J$ and $K$ into a null space.  This means that when the waveform is unknown, source localization is possible only when there exists a null space, which in turn must remain null when we get the propagation-direction right.  This property can be used for source localization \cite{tinto89} and for a consistency check for a GW from a given direction \cite{wen05a, wen07a, chatterji06}.  In addition, the value of $P_{JK}$ is directly related to the inverse of
$\hat{\mathbf A}^\dagger \hat{\mathbf A}$.  Thus if $\hat{\mathbf A}$ is singular or ill-posed in the sense
that the one of the singular values is much smaller than the other,
a pseudo-inverse should be used to calculate the correct
Fisher-matrix. In terms of implementation, this corresponds to
ignoring data corresponding to weak network
sensitivity \cite{wen07a, wen_fan, malik06}

\subsubsection{Best-case scenario: signal with known waveform but unknown arrival time}
\label{sec_best}
Now let us turn to the {\it best-case} scenario with known waveform (but unknown arrival time), we have
\begin{equation}
\delta \hat{\mathbf{d}}  = -i  e^{-i\bphi} \left[\left(\sum_j\partial_j \bphi  \delta n_j\right)   + \delta t_0 \mathbf{I}\right]\hat{\mathbf{A}}\mathbf{h}.  
\end{equation}
A similar derivation leads to 
\begin{eqnarray}
\Gamma_{jk}
& =& 
 \langle (\partial_j \bphi) \hat{\mathbf{d}}
\,| (\partial_k \bphi) \hat{\mathbf{d}}\rangle  \nonumber \\
 &-& 
 \frac{
 \langle (\partial_j \bphi) \hat{\mathbf{d}}  | \Omega \hat{\mathbf{d}}\rangle
   \langle \Omega  \hat{\mathbf{d}}|  (\partial_k\bphi)\hat{\mathbf{d}}\rangle
 }
 {\langle  \Omega \hat{\mathbf{d}} |  \Omega \hat{\mathbf{d}}\rangle}.
\label{Tau_ij}
\end{eqnarray}
After some algebraic manipulation, we obtain
\begin{eqnarray}
&&\left[\det\mbox{\boldmath$\Gamma$} \right]_{\rm best} \nonumber \\
&=&
\frac{1}{8c^4} \sum_{J,K,L,M} \frac{\xi_J  \xi_K  \xi_L  \xi_M  |(\mathbf{r}_{KJ} \times \mathbf{r}_{ML})\cdot \mathbf{n}|^2
}{\left(\sum_{I=1}^{N_d} \xi_I\right)^2}
\label{tau_best}
\end{eqnarray}
Here we have defined
\begin{equation}
\label{xiJ}
\xi_J \equiv 2 \int_{-\infty}^{+\infty}\frac{d\Omega}{2\pi} \Omega^2 |\hat{d}_J|^2.
\end{equation}
Note that $\xi_J$ is directly related to noise-weighted energy flux received by the detector (c.f. Ref.~\citep{thorne87}).

{\it Two-detector configuration in the best-case scenario: } In this case, only a 1-dimensional angular parameter can be determined.  The 1-$\sigma$ 1-dimensional angular resolution of a 2-detector network can therefore be derived directly from the trace of the Fisher matrix in Eq.~\eqref{Tau_ij}, with error angular separation defined to be twice the standard deviation $\Delta \theta_s =2/\sqrt{tr(\mbox{\boldmath$\Gamma$})}$,
\begin{equation}
\Delta \theta^{(2,\rm best)}_s = \frac{2c}{\sqrt{(\xi_1+\xi_2)} D_\perp}\sqrt{ \frac{1}{\xi_{1}\xi_{2}/(\xi_1+\xi_2)^2}},
\label{theta_2}
\end{equation}
where $D_{\perp} $ is the projected distance of the two detectors in the plane perpendicular to the wave direction, and the best determined direction is normal  to the equal time-delay lines of the two detectors.  The factor of 2 accounts for the total width of the 1-$\sigma$ angular separation.  We leave the scaling of $\xi_1, \xi_2,\cdot \cdot \cdot $ by their sum to emphasize the importance of the fractional contributions of the noise-weighted energy flux from individual detectors.

{\it Three-detector configuration in the best-case scenario:}
The simplified expression for a 3-detector network derived from Eqs.~\eqref{Omega_def}, \eqref{tau_best} and \eqref{xiJ} is 
\begin{equation}
\Delta \Omega^{(3,best)}_s = \frac{\pi c^2}{(\xi_1+\xi_2+\xi_3) A_\perp} \sqrt { \frac{1} {\xi_{1}\xi_{2}\xi_{3}/(\xi_1+\xi_2+\xi_3)^3}},
\label{ang_3det}
\end{equation} 
where $A_\perp = | (\mathbf{r}_{12} \times \mathbf{r}_{13}) \cdot \vec
{n}|/2$ is the projected area of the network perpendicular to the wave-propagation direction.  Again, the fractional contributions of the noise-weighted energy flux coupled to individual detectors play an important role in the angular resolution.

For the special case of a monochromatic GW at frequency $f$, for a 2-detector network,

\begin{equation}
\Delta \theta^{(2f,best)}_s = \frac{1}{\rho_{N}} \frac{c}{f D_\perp}\frac{1}{\pi} \sqrt{ \frac{1}{\rho^2_{1}\rho^2_{2}/\rho^4_N}}.
\label{theta_2f}
\end{equation}
For a 3-detector network,
\begin{equation}
\Delta \Omega^{(3f,best)}_s = \frac{1}{\rho^2_{N}}\frac{c^2}{f^2 A_\perp}\frac{1}{4\pi} \sqrt{\frac{1}{\rho^2_1\rho^2_2\rho^2_3/\rho^6_N}},
\label{omega_3}
\end{equation} 
where $\rho^2_I$ is the optimal SNR-squared for detector $I$ and $\rho^2_{N}$
is the optimal network SNR-squared [Eq.~\eqref{rho_SNR}].  

Compared with previous expressions summarized in \citet{sylvestre03},  Eq.~\eqref{omega_3} includes a new feature indicating that for a given network SNR, the angular resolution is limited by the least-sensitive detector. When one detector has null response to the wave (e.g., $\rho_1 \sim 0$), we have $\Delta \Omega \gg 1$, until limited by contributions from the directional derivatives of the antenna beam pattern functions we ignore in Eq.~\eqref{dd_dn}.  This is expected because if the network contained only 2 detectors then only one dimension of the source direction could be resolved. 

\subsubsection{Short signals: general expression}
\label{short_general}
Previous derivations can be extended to a general case, where in addition to angular parameters, there exists a discrete set of unknown
parameters represented in a vector $\vec\lambda$ of size $N_\lambda$. Using the same variational approach described previously, or a standard matrix inversion technique, we obtain the Fisher matrix for the angular resolution
\begin{eqnarray}
\Gamma_{ij}
&=& \langle
\partial_{i} \hat {\mathbf{d}} | \partial_{j}  \hat{\mathbf{d}} \rangle \nonumber -{\langle
\partial_{i} \hat {\mathbf{d}} | \partial_{\vec{\lambda}} \hat{\mathbf{d}} \rangle  { \langle \partial_{\vec{\lambda}} \hat{\mathbf{d}} |  \partial_{\vec{\lambda}} \hat{\mathbf{d}} \rangle}^{-1} \langle  \partial_{\vec{\lambda}} \hat{\mathbf{d}} |  \partial_{j}  \hat{\mathbf{d}}\rangle  }  \nonumber \\
 &=&\langle
\partial_i \hat {\mathbf{d}}
|\mathbf{I} - \mathbf{P}| \partial_j  \hat{\mathbf{d}}
\rangle 
\label{tau2_ij}
\end{eqnarray}
where
$ \mathbf{P}\equiv  |  \partial_{\vec{\lambda}} \hat{\mathbf{d}} \rangle {\langle \partial_{\vec{\lambda}} \hat{\mathbf{d}}  | \partial_{\vec{\lambda}} \hat{\mathbf{d}} \rangle }^{-1} \langle  \partial_{\vec{\lambda}} \hat{\mathbf{d}}|
$,  $ \partial_{\vec{\lambda}} \hat{\mathbf{d}}$ is the $N_d \times N_\lambda$ derivative matrix of the data
$\hat{\mathbf{d}}$ with respect to the unknown parameters $\vec{\lambda}$ with the matrix components defined as $(\partial_{\vec{\lambda}} \hat{\mathbf{d}})_{Ii} = \partial \hat{{d}_{I}}/ \partial_{\lambda_i}$ and the bracket operation of two matrices is defined as a matrix $\langle \mathbf{A} | \mathbf{B} \rangle_{ij} = 2\sum_I \int_{-\infty}^{+\infty} {d\Omega}/{2\pi} A^{*}_{iI}B_{Ij}$

Note that $\mathbf{P}$ is a projection operator on the vector space spanned by $\partial_{\vec{\lambda}} \hat{\mathbf{d}}$ and that $\mathbf{I}-\mathbf{P}$ is a projection operator onto the space orthogonal to that of $\partial_{\vec{\lambda}} \hat{\mathbf{d}}$.  Evidently, we have $\mathbf{P} |\partial_{\vec{\lambda}} \hat{\mathbf{d}} \rangle = |\partial_{\vec{\lambda}} \hat{\mathbf{d}} \rangle$ and $
(\mathbf{I}-\mathbf{P}) |\partial_{\vec{\lambda}} \hat{\mathbf{d}} \rangle =\mathbf{0}.
$

The matrix component of the Fisher matrix can be written from Eq.~\eqref{tau2_ij} as,
\begin{eqnarray}
\Gamma_{ij}&=& \sum_{I,J}
\Delta_{IJ}\partial_i\tau_I\partial_j\tau_J,  
\label{tau3_ij}
\end{eqnarray}
where
\begin{equation}
\Delta_{IJ} 
=\langle \Omega \hat{d}_I | \Omega \hat{d}_J \rangle \delta_{IJ} 
- \langle \Omega \hat{d}_I| \partial_{{\lambda}_k} \hat{{d}}_I\rangle 
 \mathbf{B}^{-1}_{kl}\langle \partial_{\lambda_l} \hat{{d}}_J | \Omega \hat{{d}}_J \rangle, 
\label{Delta_ij}
\end{equation}
where $\delta_{IJ}$ is a Kronecker delta
\begin{eqnarray}    
\delta_{IJ} &=& \left\{ \begin{array}{lr} 
1 & \mbox{for }\  I=J  \\ 
0 & \mbox{for }\  I \neq J\end{array} \right. ,
\end{eqnarray}
and $\mathbf{B} = \langle  \partial_{\vec \lambda}\hat{\mathbf{d}} | \partial_{\vec \lambda}\hat{\mathbf{d}} \rangle$ is a matrix with components
\begin{eqnarray}
{B}_{kl}  = 2 \int_{-\infty}^{+\infty}\frac{d\Omega}{2\pi} \partial_{\lambda_k}\hat{\mathbf{d}}^\dagger \partial_{\lambda_l}\hat{\mathbf{d}}.
\label{B_kL}
\end{eqnarray}


For a general assumption that the arrival time $t_0$ is one of the unknown parameters, $\sum_J \Delta_{IJ}=0$.  To see this, assume that $t_0$ is the $k'$-th unknown parameter.  Then $\lambda_{k^{'}}=t_0$, $\partial_{\lambda_{k'}} d_I=-i\Omega d_I$ for any $I$ and ${B}_{lk'}=-i \sum_J \langle \partial_{{\lambda_l}}\hat{{d}}_J | \Omega \hat{{d}}_J \rangle$ for any column number $l$.
\begin{eqnarray}
&&\sum_J \Delta_{IJ} \nonumber\\
&=&\sum_J \langle \Omega \hat{d}_I | \Omega \hat{d}_J \rangle \delta_{IJ} 
- \sum_{kl} \langle \Omega \hat{d}_I | \partial_{\lambda_k} \hat{d}_I \rangle 
\mathbf{B}^{-1}_{kl} \sum_J \langle \partial_{\lambda_l} \hat{{d}}_J | \Omega \hat{{d}}_J \rangle\nonumber\\
&=&\langle \Omega \hat{d}_I | \Omega \hat{d}_I \rangle 
- i\sum_{kl} \langle \Omega \hat{d}_I | \partial_{\lambda_k} \hat{d}_I \rangle 
\mathbf{B}^{-1}_{kl} B_{lk'}\nonumber\\
&=& \langle \Omega \hat{d}_I | \Omega \hat{d}_I \rangle
- \sum_{k} \langle \Omega \hat{d}_I | \partial_{\lambda_k} \hat{d}_I \rangle \delta_{kk'}\nonumber\\
&=& \langle \Omega \hat{d}_I | \Omega \hat{d}_I \rangle
- i\langle \Omega \hat{d}_I | \partial_{\lambda_{k'}} \hat{d}_I \rangle\nonumber\\
&=& 0, \nonumber
\end{eqnarray}  
where we have used $\sum_l (\mathbf{B}^{-1})_{kl}B_{lk'} =\delta_{kk^{'}}$. 
This proof enables us to obtain from Eq.~\eqref{tau3_ij}
\begin{eqnarray}
&&\left[ \det\mbox{\boldmath$\Gamma$} \right]  \nonumber \\
&=& \frac{1}{8c^4} \sum_{J,K,L,M} \Delta_{JK}\Delta_{LM}
|(\mathbf{r}_{KJ} \times \mathbf{r}_{ML})\cdot \mathbf{n}|^2, 
\label{T_geo}
\end{eqnarray} 
the same expression as Eq.~\eqref{worst_gen} in sec.~\ref{sec_worst}.

In summary, for an arbitrary network of GW detectors and arbitrary incoming GWs, the component of the Fisher matrix for angular parameters is given by Eq.~\eqref{tau3_ij}.  With the general assumption that wave arrival time $t_0$ is one of the unknown parameters,  the angular resolution has a simple geometrical form,
\begin{widetext}
\begin{eqnarray}
\Delta\Omega^{(\rm Short)}_s = \frac{4\sqrt{2}\pi c^2}{\sqrt{\sum_{J,K,L,M} \Delta_{JK}\Delta_{LM} |(\mathbf{r}_{KJ} \times \mathbf{r}_{ML})\cdot {\mathbf{n}}|^2}},
\label{Omega}
\end{eqnarray}
\end{widetext}
where $\Delta_{IJ}$ is defined in Eq.~\eqref{Delta_ij} for the general case, in Eq.~\eqref{worst} for the {\it worst-case} scenario, and it can be shown that for the {\it best-case} scenario, 
\begin{eqnarray}
\Delta_{IJ} &=&  -\frac{\xi_I \xi_J}{\sum_I \xi_I} \, \, \,\,\,\ \mbox{for}\,\,\,\, I \neq J, 
\label{best}
\end{eqnarray}
consistent with previous results in Eq.~\eqref{tau_best} and Eq.~\eqref{xiJ}. Note that only $I \neq J$ terms contribute to the angular resolution.  That is, once the waveform is known, the angular resolution does not depend on the correlation of the data between detectors.  



\subsection{Long-duration signal with known waveform}
\label{sec_long}
Finally, when a {\it long signal} with known waveform continues for a long time, it is the area mapped out by the trajectory of the detector network that determines the angular resolution.  In order to see this, 
we first note that there exist two time scales in our problem: motion of the detector network (slow), and the signal and detector response time scale (fast). We first rewrite $\delta d$ from Eq.~\eqref{d_t}
\begin{eqnarray}
\delta {d}_I(t)  &=&\left[-\delta t_0 -\sum_i  r_I^i(t) \delta n_i\right]\nonumber \\
&& \sum_{p=+,\times} f^p_{I}(t) \dot{h}_p\left[ t- t_0 -\frac{\mathbf{n}\cdot \mathbf{r}_I(t)}{c}\right] \nonumber \\
&\approx& \left[-\delta t_0 -\sum_i  \frac{r_I^i(t)}{c} \delta n_i\right] \dot{d}_I(t)
\end{eqnarray}
Here we have assumed that the antenna pattern changes at a frequency much lower than that of the signal, and that the speed of the network is much lower than the speed of light. As before, we have also ignored the change of antenna beam patterns induced by $\delta n$. We can then write 
\begin{equation}
\label{fisherlong}
\langle \delta \hat{d}_I | \delta \hat{d}_I \rangle 
=\int_0^T \dot\xi_I(t) \left[-\delta t_0 -\sum_i  \frac{r_I^i(t)}{c} \delta n_i\right]^2 dt  
\end{equation}

where 
\begin{equation}
\dot{\xi}_I(t) \equiv 2 \int  \dot{d}_I(t-\tau/2) w_I(\tau)\dot{d}_I (t+\tau/2)d\tau\,,
\end{equation}
where $w_I$ is the inverse Fourier transform of $1/S_I(\Omega)$.  Note that $\dot{\xi_I}$ can be viewed as the rate of increase of the signal-to-noise ratio at time $t$ --- when we regard $\dot{d}$ as signal, with the original noise spectrum. 
In obtaining  Eq.~\eqref{fisherlong}, we have assumed the observation time, as well as the time scale at which $\mathbf{r}_I(t)$ changes,  to be much longer than the signal correlation time (i.e., the range of integration for $\tau$).  Following the same variational approach described previously, the Fisher matrix can be obtained by minimizing Eq.~\eqref{fisherlong} over $\delta t_0$, we obtain
\begin{equation}
\Gamma_{jk} = \frac{1}{c^2}\left[\overline{r_j r_k} -\bar{r}_j\,\bar{r}_k\right]\sum_J \xi_J
\end{equation}
where $\xi_J$ is defined in Eq.~\eqref{xiJ},
\begin{eqnarray}
\label{exp}
\overline{r_k} &\equiv& \frac{\sum_J \int_0^T r_J^k(t) \dot{\xi}_J(t) dt}{\sum_J \xi_J(T)}\\
\label{var}
\overline{r_j r_k} &\equiv& \frac{\sum_J \int_0^T r_J^j(t)  r_J^k(t) \dot{\xi}_J(t) dt}{\sum_J \xi_J(T)}
\end{eqnarray}
are the {\it average} of the $k$-th coordinate of detectors in the network, throughout detection time, and {\it correlations} between the $j$-th and $k$-th coordinates.  When the detector trajectory has a size much bigger than that of the array (as in the case of observing GWs from pulsars using the existing LIGO-Virgo network), we can omit the $J$ index of $r_J^k$, simply replacing it by the mean position of the detector network $r^k$, and the weighted average is simply over time.  The determinant of the inverse of the Fisher matrix yields the angular resolution, 
\begin{equation}
\Delta\Omega^{(\rm Cont)}_s = \frac{2\pi c^2}{\sum_J \xi_J(T) \sqrt{\mathrm{Var}(r_\theta)\mathrm{Var}(r_\phi) -\mathrm{Cov}^2(r_\theta,r_\phi)}},
\label{long_general}
\end{equation}
where $\theta$,
$\phi$ represent the longitude and latitude like coordinates (e.g., right ascension and declination in the celestial coordinate system).  For a general polar coordinate system, $r_\theta= \partial_\theta \mathbf{n}\cdot \mathbf{r}$ and that $r_\phi = \partial_\phi \mathbf{n}\cdot \mathbf{r}$.   Var and Cov are variance and co-variance with expectation values calculated according to Eqs.~\eqref{exp} and \eqref{var}. The angular resolution is now clearly related to the square of the area mapped out by the trajectory of the network.

{\it Detectors on circular orbits:} As a simple example, we consider the situation where the detector network makes a circular motion with radius $R_*$ and angular frequency $\omega_*$.   Supposing that $\dot{\xi}_J$ are all constants, then in the situation where the network trajectory has mapped out a size much larger than the size of the network, yet $\omega_*T \ll 1$, we find
\begin{equation}
\left[\mathrm{Var}(r_\theta)\mathrm{Var}(r_\phi) -\mathrm{Cov}^2(r_\theta,r_\phi)\right] 
= \frac{(\omega_* T)^6}{8640}R_*^4\sin i^2_{n},
\end{equation}
where $i_{n}$ is the angle between the source direction and plane of the circular motion. The factor $\sin^2 i_{n}$ accounts for the projected area from by the detector network's trajectory. Therefore, for a short observation, 
\begin{equation}
\Delta \Omega^{(\rm Cont, S)}_s = \frac{12\sqrt{15} c^2}{f^2\rho^2_{T} \pi R^2_* (\omega_* T)^3|\sin i_{n}|},
\label{pulsar_short}
\end{equation}
where $\rho_T$ is the optimal signal-to-noise ratio within the observation time $T$.  In this case, the error
area decreases like $T^{-4}$ (c.f. Refs.~\cite{schutz89,
  prix07}), where $T^{-1}$ comes from the increment of SNRs with time
and $(\omega_* T)^{-3}$ from the increment of the area formed by the
trajectory of the entire detector network. For longer observations of $\omega_* T \gg 1$, contributions from the area saturate because the maximum area is $\sim \pi R^2_*$,  
\begin{equation}
\Delta \Omega^{(\rm Cont, L)}_s = \frac{c^2}{f^2\rho^2_{T} \pi R^2_* |\sin i_{n}|}.
\label{pulsar_long}
\end{equation}
That is, the error area decreases like $T^{-1}$ as a result of the increment of total SNR-squared. Multiple detectors can be treated in the same way as a single detector, since the area formed among several detectors is much less than the area formed by the detector's trajectory over the typical observation time of much longer than minutes.  Multiple detectors thus contribute mainly by increasing SNR.  

\section{Implications}
\label{implication}`

We emphasize the clear geometrical meaning of $ |(\mathbf{r}_{KJ} \times
\mathbf{r}_{ML})\cdot {\mathbf{n}}|$ in Eq.~\eqref{Omega} for {\it short signals}, which is twice the area formed by the projections of detectors $J$, $K$, $L$
and $M$ onto the plane orthogonal to the wave propagation
direction.  This means that, when all detectors from the network form a plane, the angular resolution is better for sources at directions perpendicular to the plane but is poor for directions along the plane. A similar geometrical term is shown in Eq.~\eqref{long_general} for {\it long signals} of known waveforms.

Our formula is consistent with the understanding that a larger network improves angular resolution.  As we show in sec. V, addition of either LCGT or AIGO to the current ground-based LIGO-Virgo detector network results in a significant improvement of the angular resolution for short signals.  Inclusion of the southern hemisphere detector AIGO will improve dramatically the angular resolution of the network and also break the degeneracy in the angular resolution along the plane formed by detectors on the northern hemisphere \cite{aigo,blair07}. For observations of {\it long signals} where the size of the network is less important than the area mapped out by the trajectory of the network, long observing time is essential. 

For {\it short signals}, besides geometrical area,  the angular resolution of a detector network can be
improved by maximizing values of $\Delta_{IJ}$ in Eq.~\eqref{Omega}.   In the {\it worst-case} scenario where the
waveform is unknown, the angular resolution largely depends on the noise-weighted correlation of responses between detectors (Eq.~\eqref{worst}).  In the {\it best-case} scenario, when wave parameters except for arrival time are known, the best achievable angular resolution strongly depends on the fractional energy flux coupled to individual detectors and does not depend on correlation of data between detectors.  Therefore, building detectors of parallel antennae and comparable sensitivity is advantageous for localizing GWs in either improving the data correlation ({\it worst case}) or balancing fractional energy fluxes among detectors ({\it best case}).  

In the {\it best-case} scenario, the angular resolution is approximately inversely proportional to the total received GW energy flux weighted by noise.   When the noise spectral density $S_I(f)$ can be replaced with a characteristic value, $S_I(f) \sim S_{0}$, it follows from Eq.~\eqref{ang_3det} that for a 3-detector network, 
\begin{equation}
\Delta \Omega^{(3)}_s \propto \frac{S_0}{ \Delta E_{\rm GW}},
\label{ang_eng}
\end{equation}
where
$\Delta E_{\rm GW} = A^{*}_\perp \sum^{N_d}_{I=1}\frac{\pi c^3}{2G} \int^\infty_0 df f^2 |d_I(f)|^2$ is the GW energy coupled by the network (c.f. Ref.~\citep{thorne87}).  For GWs from coalescing binaries of two neutron stars (inspiral source),  even though the mergers occur around 1.5 kHz, the noise-weighted energy flux peaks around 150 Hz as this is where the LIGO detectors are the most sensitive.   This makes the inspiral GW sources relatively low-frequency sources in terms of localization.  

For a given detector network of fixed configuration, the same considerations should be given for localization strategies.  For sources of known waveforms, we need to focus on maximization of the total coupled energy flux weighted by noise while maintaining a balanced budget for fractional energy flux coupled to individual detectors.  For localizing GWs of unknown waveforms, it is crucial to maximize the noise-weighted data correlation between detectors and to make use of the maximum dimension of the null-space~\cite{wen07a, wen_fan, searle09, malik06}. 

\section{Astrophysical Applications}
\label{astro}
We first apply our formulae to the existing detector network, that is, the 3-detector network consisting of the LIGO detectors at Livingston (L) and Hanford (H) in the US and Virgo (V) in Italy~(the network is often called the LHV network) with 4 km-armlengths.   When there are only two detectors available at different sites, say the two LIGO detectors (L and H), we can locate the source with 1-dimension angular resolution using Eq.~\eqref{theta_2f}. The angular separation of the 1-$\sigma$ error-bars is
\begin{equation}
\Delta \theta^{(\rm Short, 2f)} \approx  5^\circ \frac{150{\rm Hz}}{f} \frac{10}{\rho_N} \frac{ 3000 {\rm km} }{D} \sqrt{\frac{(1/2)^2}{\rho^2_1\rho^2_2/\rho^4_N}}\frac{\sqrt{2}/2} {|\sin i_{n}|}, 
\label{ang_2f}
\end{equation}
where $D$ is the distance between the two detectors,  $i_{n}$ is the angle between the 2-detector baseline and wave direction.  The error bar is about a factor of 3 better for the LIGO-Virgo baseline.  For a 3-detector network, 
\begin{widetext}
\begin{equation}
\Delta \Omega^{(\rm Short, 3f)}_s \approx 8 \ \mbox{sq-degs} \left ( \frac{150{\rm Hz}}{f} \frac{10}{\rho_N} \right )^2 { \frac{ 10^{17} {\rm cm^2} }{A_N} \frac{1/27}{\rho^2_1\rho^2_2\rho^2_3/\rho^6_N} \frac{\sqrt{2}/2} {|\sin i_{n}|}},
\label{ang_3f}
\end{equation}
\end{widetext}
where $A_N$ is the triangular area formed by the three detector sites in the detector network. We adopted $A_N=10^{17}$cm$^2$ for the LIGO-Virgo network. $i_{n}$ is the angle between the wave direction and the plane formed by the three detectors. 

We also apply Eq.~\eqref{pulsar_short} to the ground-based observations of monochromatic GWs, e.g.,  from pulsars.   For a typical observation time longer than minutes, the angular resolution is determined by the area of the virtual network formed by the detectors' Earth-Sun motion which dominates over the area formed by Earth's self-rotation. The ``{\it best case}'' scenario for observations of minutes $\ll T \ll$ months is
\begin{widetext}
\begin{equation} \Delta \Omega^{(\rm Cont, S)}_s \approx 2.4 \ \mbox{sq-arcmins}   \left ( \frac{900 {\rm Hz}} {f} \frac {10}{\rho_T}\right )^2 {\left (\frac{1  \mbox{d}}{T}\right )^3\frac{\sqrt{2}/2}{|\sin i_{n}|}}. \end{equation}\end{widetext}
For longer observations, our approximation of monochromatic wave is most likely no longer valid as higher-order derivatives of frequencies generally must be considered (e.g. Ref.~\cite{schutz89}). 

The angular resolution for a monochromatic source with detectors on the Earth-Sun orbit can also be written using Eq.~\eqref{pulsar_long} 
for long observations, e.g., more than a year's observation using the future space detector LISA at its low frequency range,
\begin{equation}
\Delta \Omega^{(\rm Cont, L)}_s \approx 0.73 \ \mbox{sq-degs}  \left ( \frac{3 \rm {mHz}} {f} \frac {30}{\rho_T}\right )^2 {\frac{\sqrt{2}/2}{|\sin i_{n}|}}. 
\end{equation}
Our result is consistent with a simple estimation using the diffraction limit \cite{schutz09} but is a factor of a few smaller than the results in Ref.~\citep{cutler98} probably because our result represents the {\it best-case} scenario where we assume perfectly known waveforms.

\begin{figure}
\includegraphics[clip, width=3.0in]{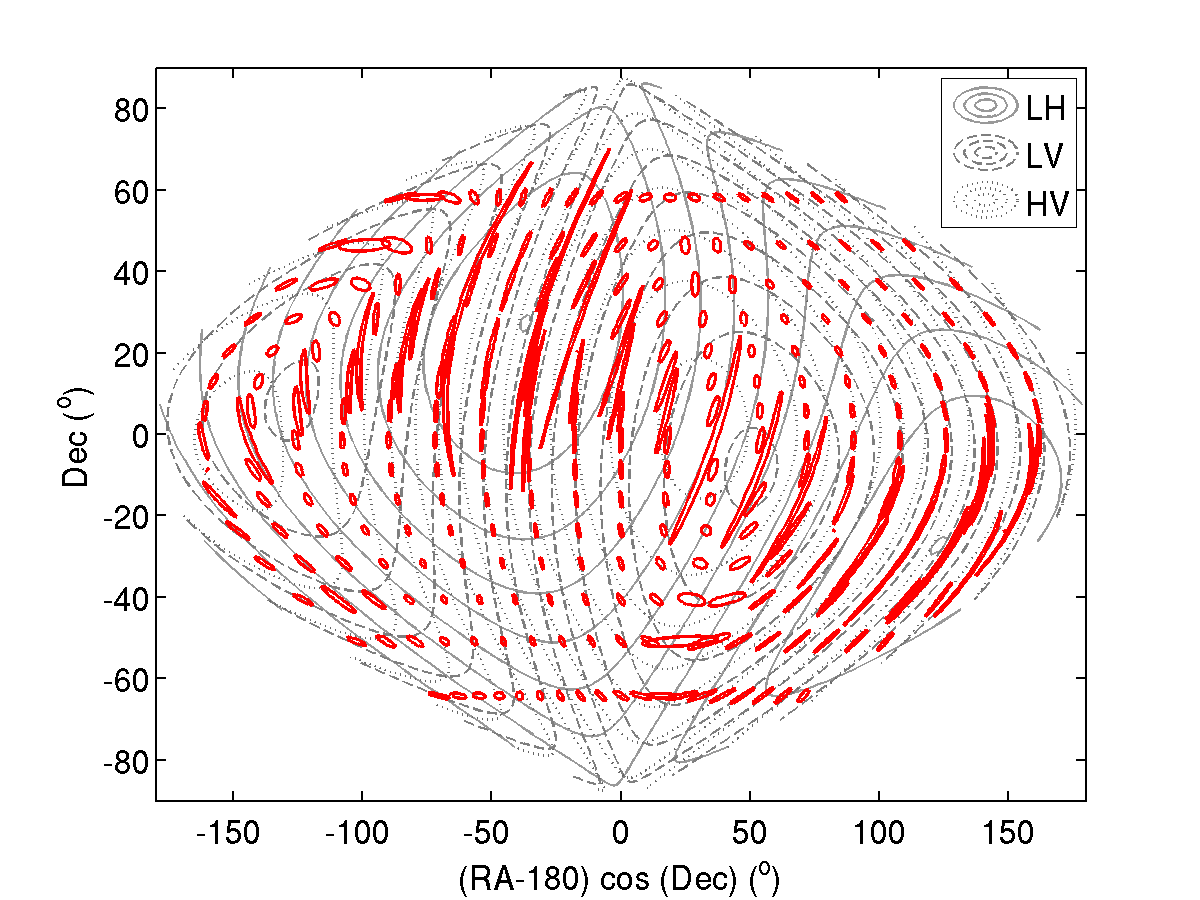}
\caption{All-sky map of error ellipses of angular parameters at the 95\% confidence calculated from Eq.~\eqref{Tau_ij} for the 3-detector network of the Advanced LIGO detectors (L and H) and the Advanced Virgo (V) to detect GWs from bar-instability that center around 600 Hz for the {\it best-case} scenario [see also Eqs.~\eqref{Omega},\eqref{best}].  These ellipses were calculated at a fixed time for a fixed optimal network SNR $\rho_N=10$.  Shown also in the background are contours of light arrival-time delays between detector pairs at a 2 ms interval for the L-H pair and 4 ms intervals for all other pairs.   
}\label{all_sky_LHV}
\end{figure}

\begin{figure}
\includegraphics[clip, width=3.0in]{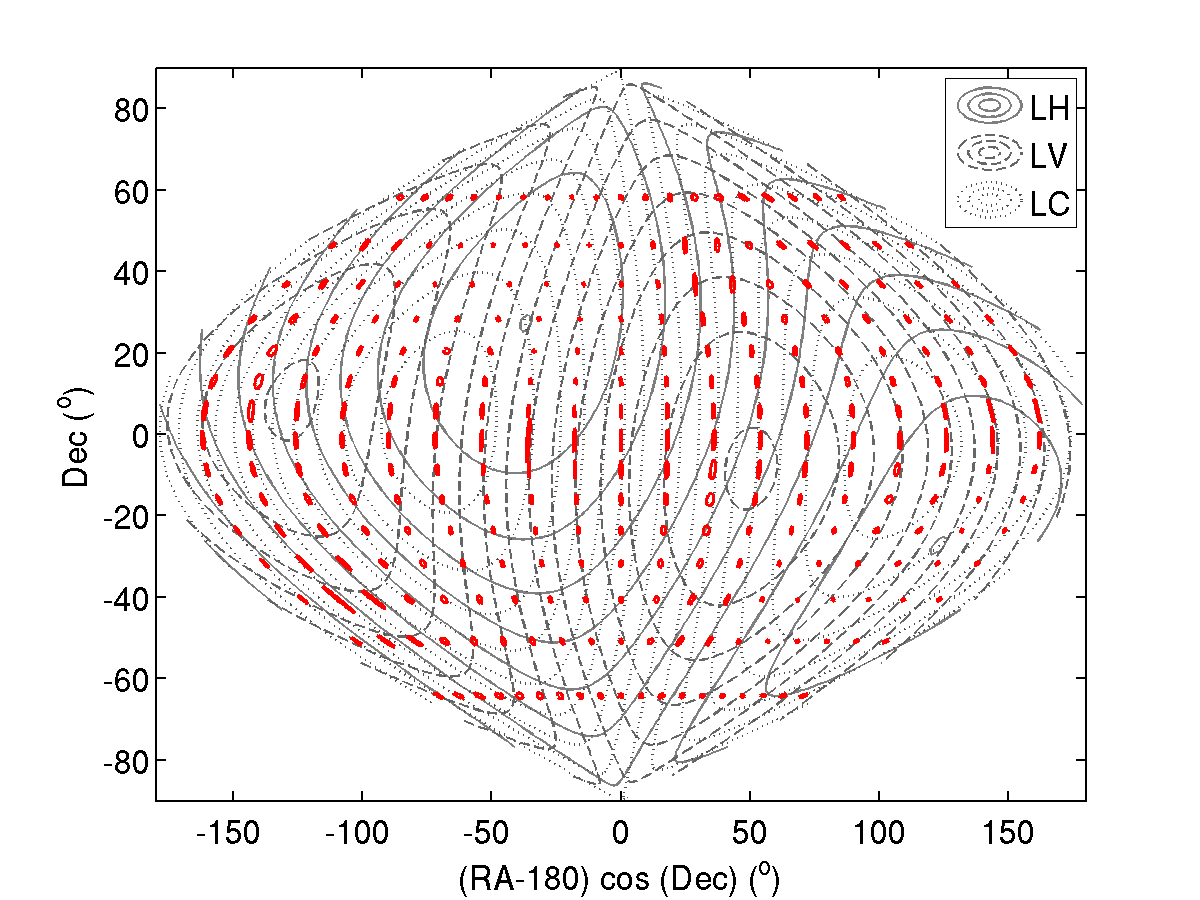}
\caption{Same as Fig.~\ref{all_sky_LHV} but with LCGT (C) in Japan added to the LIGO-Virgo network.  For LCGT, the Advanced-LIGO sensitivity is assumed. 
}\label{all_sky_LHVC}
\end{figure}

\begin{figure}
\includegraphics[clip, width=3.0in]{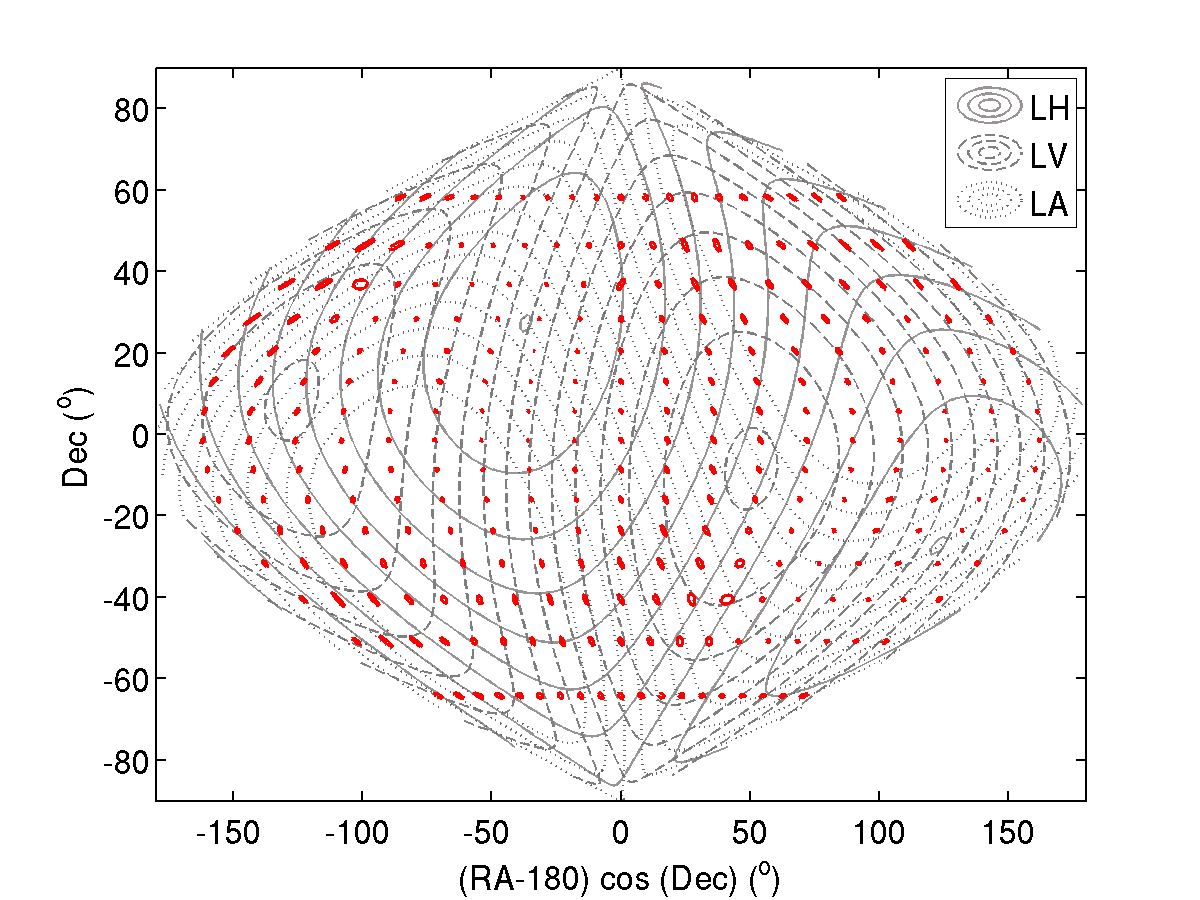}
\caption{Same as Fig.~\ref{all_sky_LHV} but with AIGO (A) from Australia added to the LIGO-Virgo network.  For AIGO, the Advanced-LIGO sensitivity is assumed. 
}\label{all_sky_LHVA}
\end{figure}

\begin{figure}
\includegraphics[clip, width=3.0in]{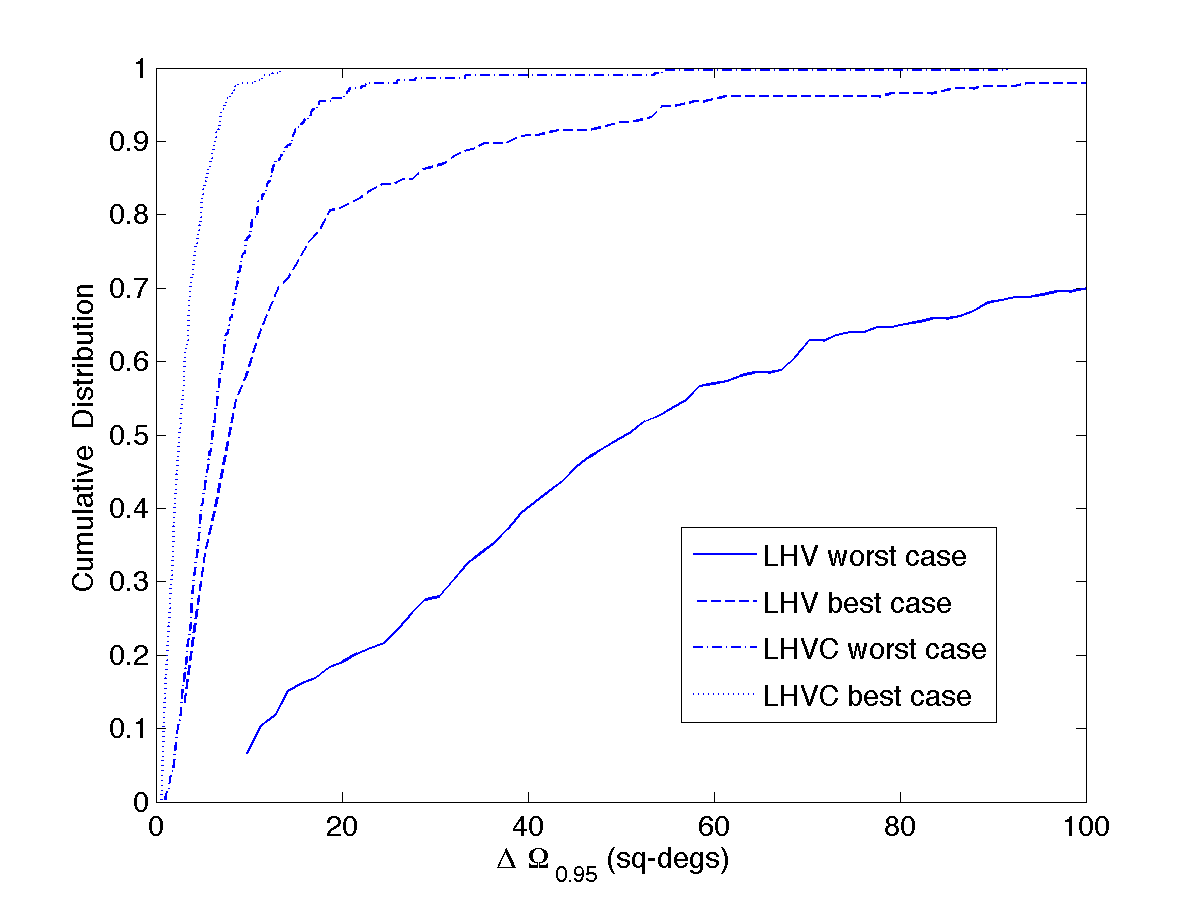}
\caption{Cumulative distribution of areas of error ellipses for angular parameters at 95\% confidence, $\Delta \Omega_{0.95}$, for the {\it best-case} (shown in Fig.~\ref{all_sky_LHV} and in Fig.~\ref{all_sky_LHVC}) and corresponding {\it worst-case} scenario for the LIGO-Virgo (LHV) and LIGO-Virgo-LCGT (LHVC) network.        
}\label{LHVC_cum}
\end{figure}

\begin{figure}
\includegraphics[clip, width=3.0in]{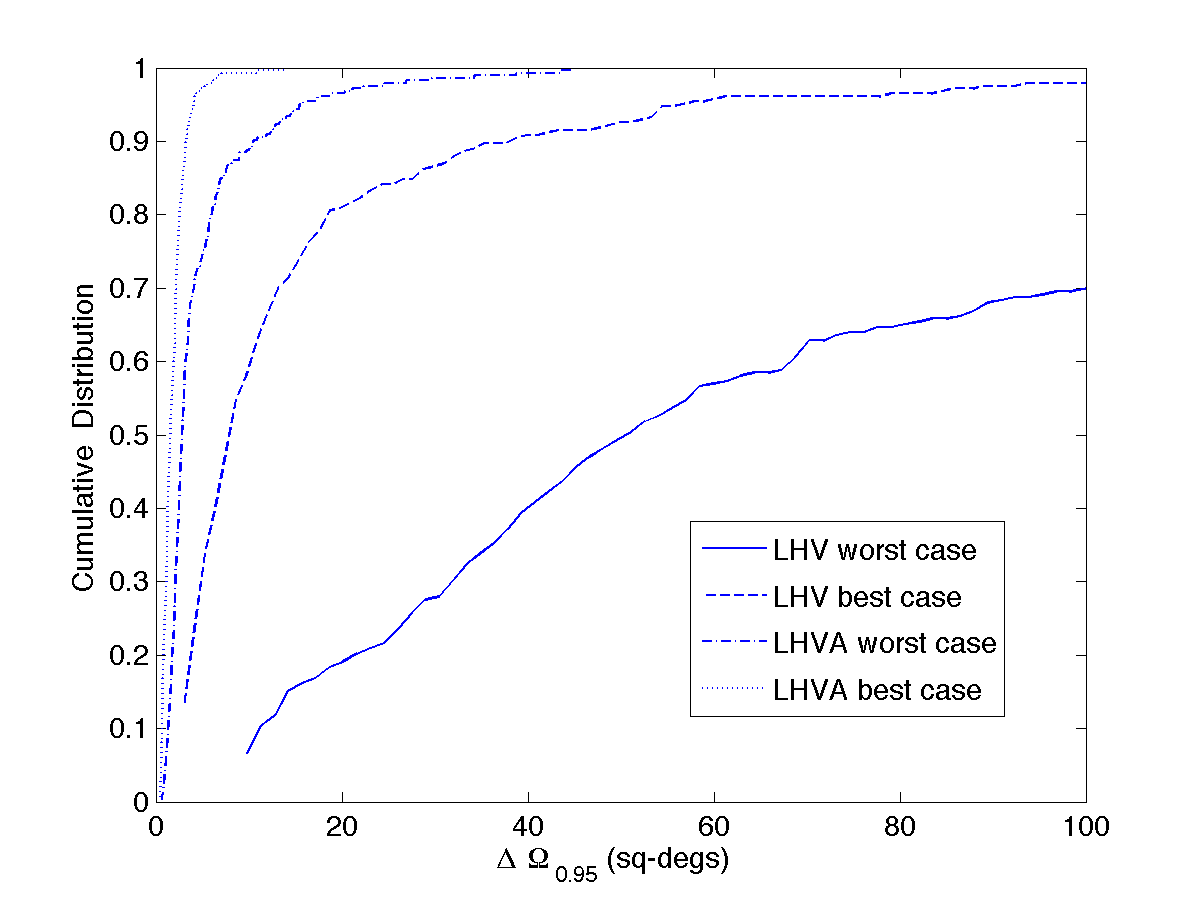}
\caption{Same as Fig.~\ref{LHVC_cum}, but with AIGO added as the 4th detector (denoted the LHVA network) instead of LCGT. 
}\label{LHVA_cum}
\end{figure}

\begin{figure}
\includegraphics[clip, width=3.0in]{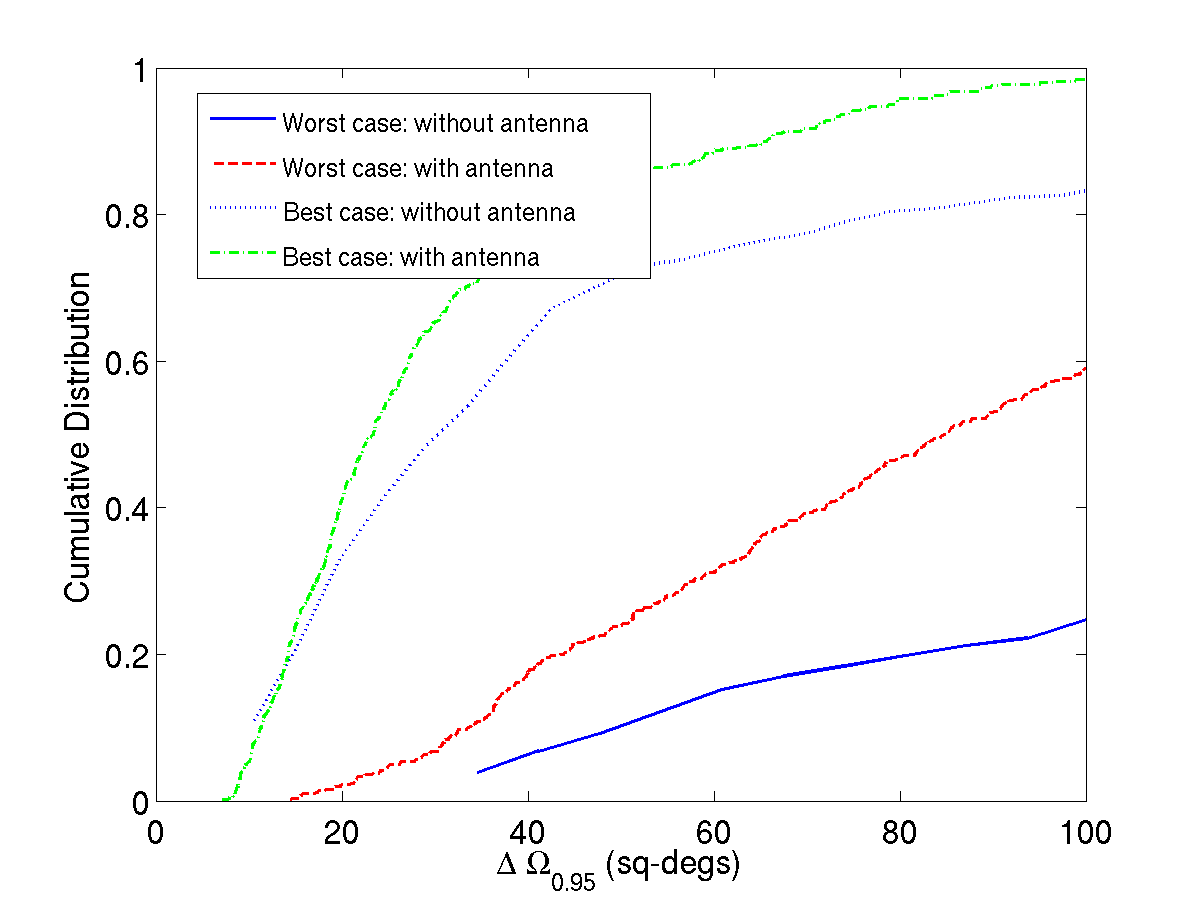}
\caption{Cumulative distribution of the areas $\Delta\Omega_{0.95}$ of error ellipses at 95\% confidence for the {\it best}- and for the {\it worst-case} scenario for GWs from coalescing NS-NS binaries at the optimal network SNR $\rho_N=15$ at random times and random orientations of the binaries (see text).   We use the 3-detector network of the initial LIGO detectors (L and H) and initial Virgo (V) at the design sensitivity. 
}\label{all_sky_nsns_cum}
\end{figure}

\begin{figure}
\includegraphics[clip, width=3.0in]{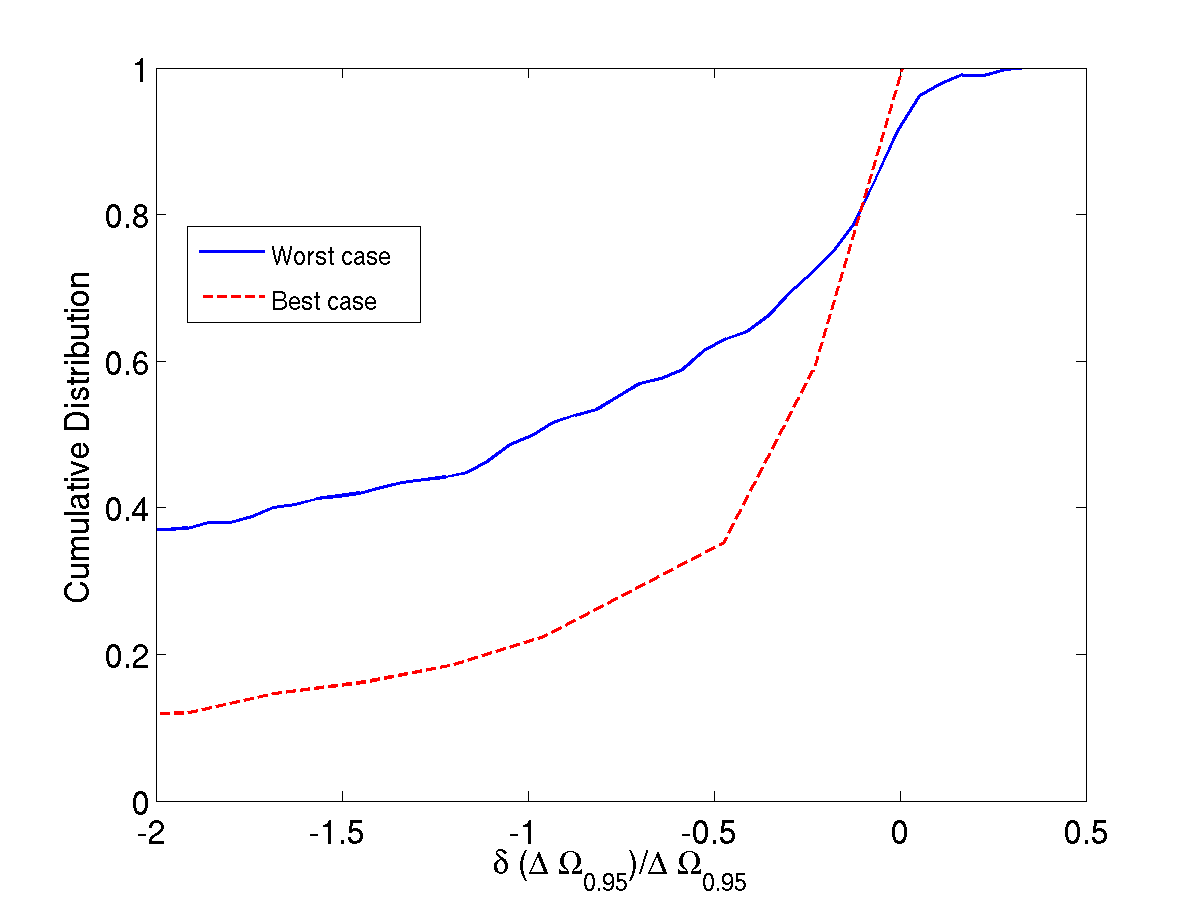}
\caption{Cumulative distribution of fractional difference of the angular resolution for GWs from coalescing binaries of neutron stars with and without contributions from direction derivatives of antenna beam pattern functions.  Negative values in X-axis correspond to overestimation of $\Delta \Omega_{0.95}$.  Same data for Fig. \ref{all_sky_nsns_cum} are used. 
}\label{insp_ant}
\end{figure}

We show in Figs. \ref{all_sky_LHV}--\ref{insp_ant} sky maps and statistical behavior of the error ellipse of angular parameters derived for {\it short signals} in Eqs~\eqref{Omega} for the existing three detector network of LIGO and Virgo and for a possible 4-detector network in the future.  With the assumption of stationary Gaussian noise, at 95\% confidence level, the area of the error ellipse
\begin{equation}
\Delta \Omega_{0.95} \approx 2.5 \Delta \Omega_s.  
\label{O_0.95}
\end{equation}
Fig.~\ref{all_sky_LHV} shows an all-sky map of the error ellipses in the {\it best-case} scenario calculated from Eq.~\eqref{Tau_ij} for the LIGO-Virgo 3-detector network at the advanced configuration to detect model GWs from the bar-mode instability of supernova (the U11 model as it is from Ref.~\cite{mario}). The duration of the wave is about 20 ms with central frequency around $600$ Hz.   Similar results are obtained if we use more sophisticated waveforms from Ref.~\cite{ott}.  The GWs are injected uniformly in 270 sky directions at a given time.  The distances to the source are adjusted so that the 3-detector optimal network signal-to-noise ratio $\rho_N=10$.  Fig.~\ref{all_sky_LHVC} and Fig.~\ref{all_sky_LHVA} show the same all-sky maps if the planned GW detector  LCGT (C) or AIGO (A) is added.  The noise spectral density of all detectors are drawn from the design sensitivity for the Advanced LIGO detectors \cite{sf_aligo} (with zero-detuning of the signal recycling mirror and high laser power). 

It is apparent that for the {\it best-case} scenario, the shapes of error ellipses are determined by contours of equal light arrival-time delays between detectors.   With addition of a new detector to the 3-detector network, there is a significant improvement to the angular resolution due to longer baselines and improved SNRs.  Fig.~\ref{all_sky_LHV} shows that when all detectors are in the northern hemisphere, the angular resolution is relatively poor along the plane formed by these detectors.  Compared to LCGT,  the location of AIGO in the southern hemisphere more efficiently lifts the plane-degeneracy formed by detectors in the northern hemisphere.

Fig.~\ref{LHVC_cum} and Fig.~\ref{LHVA_cum} show cumulative distributions of $\Delta \Omega_{0.95}$, the areas of error ellipses shown in Figs.~\ref{all_sky_LHV}--\ref{all_sky_LHVA}.  The improvement with a larger network at the {\it worst-case} scenario is more prominent than the {\it best-case} one because, in addition to improvement in SNRs and geometrical area,  the dimension of the null-space doubles when the fourth detector is added (sec~\ref{sec_worst}).  Our results show that for the advanced configuration, 50\% of the sources with SNR $\rho_N=10$ can be best localized within 8 sq-degs for the {\it best cases} and within 50 sq-degs for the {\it worst cases} for the LIGO-Virgo network.  These areas will be reduced by a factor of 2.5 and 10 for the {\it best-case} and the {\it worst-case} scenarios respectively if LCGT is added to the LIGO-Virgo network (Fig.~\ref{LHVC_cum}).  Similarly, factors of 6 and 15 reduction are observed for these error areas for the {\it best-case} and the {\it worst-case} scenarios respectively if AIGO is added instead (Fig.~\ref{LHVA_cum}) .   


\begin{figure}
\includegraphics[clip, width=3.0in]{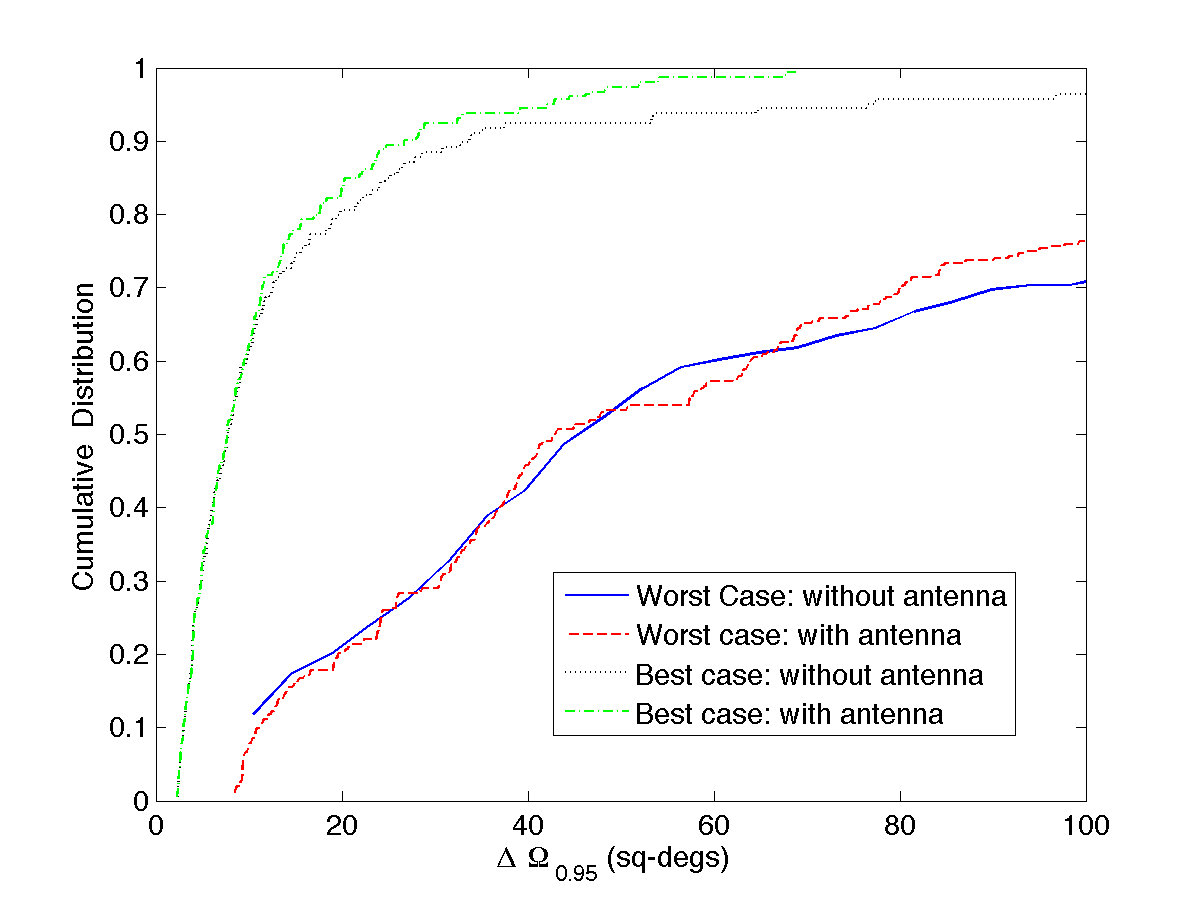}
\caption{Comparison of cumulative distributions of $\Delta \Omega_{0.95}$ for the LIGO-Virgo network for GWs from the bar-mode instability with and without contributions from direction derivatives of antenna beam pattern functions. 
}\label{SN_ant}
\end{figure}

\begin{figure}
\includegraphics[clip, width=3.0in]{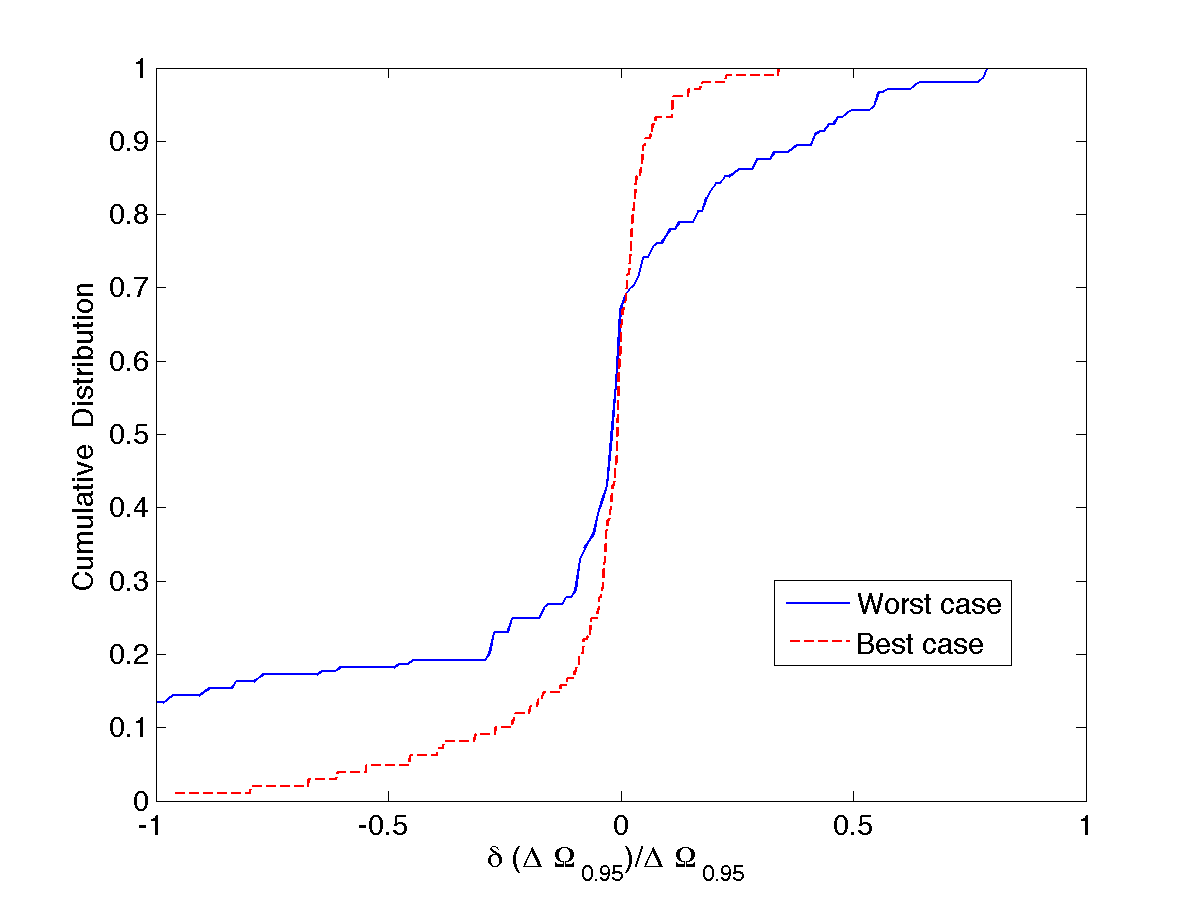}
\caption{Cumulative distribution of fractional difference in the angular resolution for GWs of the bar-mode instability caused by ignoring direction derivatives of antenna beam pattern functions.  The same data as in Fig. \ref{SN_ant} are used. 
}\label{SN_ant_err}
\end{figure}


For the interesting case of GWs from coalescing binaries of two neutron stars, we show in Fig.~\ref{all_sky_nsns_cum} the cumulative distribution of the solid angle $\Delta\Omega_{0.95}$ of the error ellipse for the network consisting of the initial LIGO (LH) detectors and initial Virgo with the design sensitivity \cite{sf_iligo,sf_virgo}.  We adopt the conventional second-order post-Newtonian approximation for the phase of GW waveforms and Newtonian amplitude \citep{pe, lal}.  The masses of the neutron stars are chosen to be 1.4\ \msun\, each.   The distances to the source are again adjusted so that the optimal network SNR is $\rho_N=15$.  The GW signals are injected randomly in 500 sky directions at randomly chosen times within a sidereal day and with polarization angles as well as binary inclinations drawn from uniform distributions.  

It is apparent in Fig.~\ref{all_sky_nsns_cum} that 50\% of the sources detected by this initial LIGO-Virgo network have $\Delta \Omega_{0.95} \le 23$ sq-degs ({\it best case}) and $\Delta \Omega_{0.95} \le 80$ sq-degs for the {\it worst case}.  This and results for higher frequency burst-like GWs discussed in Fig.~\ref{LHVC_cum} and Fig.~\ref{LHVA_cum} mean that wide-field EM cameras with reasonable sensitivity and angular resolution are advantageous for follow-up observations to best catch the electromagnetic counterparts of GW sources with current LIGO-Virgo network.  We have also found similar cumulative distributions for the advanced detectors for the same fixed SNRs. Within such large error ellipses, if improved localization (e.g., by X-ray counterparts) is not available, careful strategies need to be designed to identify host galaxies using wide-field optical or radio telescopes \citep{blair07}.  A detailed study in this aspect using the most up-to-date galaxy catalog and applying the cataloged B-band luminosity to trace possible GW events is on-going \cite{phinney09}. The addition of LCGT or AIGO can bring a factor of $\ge 10$ reduction for the medium values of $\Delta \Omega_{0.95}$ in the {\it worst-case} scenario and a factor of a few for the {\it best-case} scenario, and will help significantly the pointing of EM telescopes for follow-up observations and  elimination of confusion sources.
\section{Contributions of Antenna Beam Patterns} 
\label{sec_antenna}

Our derivation of geometrical expressions for the angular resolution of a network of GW detectors does not take into account the directional derivatives of antenna beam pattern functions (Eq.~\eqref{dd_dn}).  It is foreseeable that such derivatives can be important for low-frequency signals if $|\Omega \partial_\theta \tau| \le |\partial_\theta \log f^{+,\times}|$ is approximately true,  where $\tau$ is time-delay between pairs of detectors. On the other hand,  this additional effect will mainly lower the lower-limits on $\Delta \Omega$ calculated in this paper, meaning, most of the $\Delta \Omega_s$ or $\Delta \Omega_{0.95}$ given here are over-estimates of their true values.  We define the fraction changes in $\Delta\Omega_{0.95}$ as
\begin{equation}
\delta (\Delta\Omega_{0.95})/\Delta\Omega_{0.95} = 1-\Delta \Omega_{0.95}/\Delta \Omega^{A}_{0.95}, \ \
\end{equation}
where $\Delta \Omega_{0.95}$ is the error area we calculate in this paper, $\Delta\Omega^A_{0.95}$ is the actual value if the derivatives of antenna beam pattern functions with respect to sky directions are taken into account in Eq.~\eqref{dd_dn}.  

The inspiral sources, one of the most important GW sources for the ground-based detectors, are considered to be ``low-frequency'' sources as discussed previously with the noise-weighted energy flux peaked around 150 Hz for a 1.4-1.4 \msun\ binary GW source for the initial or Advanced LIGO.   The effect of the directional derivatives of antenna beam pattern functions is therefore not completely negligible compared to the time-delay effect.  We found that (Fig.~\ref{all_sky_nsns_cum}) the cumulative distributions with and without such effect are consistent with each other within 22\% for $\Delta \Omega_{0.95} \leq 23$ sq-degs in the {\it best-case} scenario. The discrepancy is significantly larger ($\ge 60$\% ) for the {\it worst-case} scenario where $\Delta \Omega_{0.95} \ge 35$.  That is, the effect on the cumulative distribution of $\Delta \Omega_{0.95}$ is significantly larger for large values of $\Delta \Omega_{0.95}$ and for the {\it worst cases}. On the other hand,  Fig.~\ref{insp_ant} shows that,  in the {\it best-case} scenario, about 35\% of $\Delta \Omega_{0.95}$ are overestimated by more than 50\%.    For the {\it worst-case} scenario,  about 62\% of $\Delta \Omega_{0.95}$ are overestimated by more than 50\%.   Fig.~\ref{insp_ant} also shows that a much smaller fraction of $\Delta \Omega_{0.95}$ are underestimated.  


The effect of the directional-derivatives of antenna beam pattern functions for GWs at higher frequencies is demonstrated for the supernova bar-instability case in Fig.~\ref{SN_ant} and Fig.~\ref{SN_ant_err}.  In this case, the central frequency is about $600$ Hz.  As a result of higher frequencies, compared to the inspiral sources, the difference in cumulative distributions of $\Delta \Omega_{0.95}$ is less significant (Fig.~\ref{SN_ant}).  The difference is less than a few percent for the {\it best-case} scenario and less than 10\% for the {\it worst case} scenario for $\Delta \Omega_{0.95} \leq 100$ sq-degs.  Fig.~\ref{SN_ant_err} shows that 15\% of $\Delta \Omega_{0.95}$ are overestimated by more than 30\% for the {\it best-case} scenario, while 25\% are overestimated by the same amount for the {\it worst-case} scenario.  The fraction of $\Delta \Omega_{0.95}$  underestimated are higher than the inspiral sources though.  



While directional-derivatives of the antenna beam pattern functions contribute to changes of the areas of error ellipse, they do not significantly alter the direction of the axes.  A histogram using the same data for the inspiral case shows that 99.6\% of the time, the changes of the angles are within $\pm 10$\%.  For the example of the burst-like case, 98\% of the time, the fractional changes of the angles of axes are less than $\pm 5$\%.

\section{Conclusions}
\label{concl}

In summary, we have derived for the first time analytical expressions for the angular resolution for an arbitrary  network of gravitational-wave detectors where the directional derivatives of antenna-beam pattern functions are ignored.   Our results demonstrate the explicit dependence of the angular resolution on the geometrical configuration of the network,  the total noise-weighted energy flux coupled to the network, its fractional distribution to individual detectors, and correlation of data between detectors. These dependences are intrinsic to the configuration of the detector network and the sources.  The results are method-independent and they correspond to the best possible localization any unbiased methods can achieve in the presence of statistical errors.  Our estimate is conservative especially for low-frequency sources as the directional derivatives of antenna beam pattern functions are ignored (see sec.~\ref{sec_antenna}).  In reality, careful designs, e.g., to remove multiple local minima or maxima caused by interference when combining data from different detectors, or to break the mirror degeneracy in arrival time delays for the three-detector case (by using the wave amplitude information),   are required for localization methods in order to achieve the ``best'' limits. 

Derivation of the angular resolution for {\it short signals} including those for a network of two and three detectors are presented in Eq.~\eqref{theta_2} and Eq.~\eqref{ang_3det}.  Our results are consistent with what was previously known from the diffraction limit: that a larger network yields better angular resolution.  We confirm that the angular resolution is poor along the plane formed by current LIGO-Virgo detectors and is better for directions normal to the plane.  Numerical results are included to show how a new detector in Japan (LCGT) or in Australia (AIGO) can dramatically improve the angular resolution of the existing network (Fig.~\ref{LHVC_cum} and Fig.~\ref{LHVA_cum} and discussions thereafter), by contributing to longer baselines, additional energy flux, extended null signal space, and by breaking the plane-degeneracy formed by three detectors.   

Compared with previous approximate expressions for 3-detectors (sec.~\ref{intro}), our results are more rigorous and show the explicit roles of individual detectors.  The angular resolution of a detector network depends on the fractional energy flux coupled to individual detectors in the {\it best case} and on correlation of data between detectors in the {\it worst case}.   In the {\it best-case} scenario, the angular resolution is largely limited by the least sensitive detector.   These have significant implications for the design of a detector network and for the design of an optimal localization method for a fixed network (see sec.~\ref{implication} for more discussion on implications).  Moreover, our results apply to an arbitrary network of any number of detectors. 

We have also derived a geometrical expression for {\it long signals} for a simplified case where waveforms are known (Eq.~\eqref{long_general}).  The situation where detectors are in circular motion and the signal is monochromatic is discussed.  The dependence of the angular resolution on the latitude of the source is apparent in our formulae.   We also demonstrate that the angular resolution improves rapidly with the observing time $\sim T^4$ initially with short observations and saturates to $\sim T$ for longer observations (Eq.~\eqref{pulsar_short}, Eq.~\eqref{pulsar_long}), consistent with previous knowledge~\cite{schutz89, prix07}.  

We have further presented through numerical simulations the distribution of the areas of error ellipses at 95\% confidence level for two of the most important GW sources for ground-based detectors. (1) We apply our calculations to GWs from coalescing binaries of neutron stars using the sensitivity curves of detectors that are operating at this writing.  The actual limit of the angular resolution for these inspiral sources should be closer to the {\it best-case} scenario since theoretical waveforms are known and essential parameters can be estimated with great accuracy independent of source direction determination \cite{cutler94}.  (2) We apply our method to burst-like GWs using a representing waveform from bar-instability of neutron stars in supernovae for advanced detectors.  The actual angular resolution for this type of ``burst'' source fits in our {\it worst-case} scenario for {\it short signals}.  We shows that, for the existing LIGO-Virgo detector network, assuming uniform distribution of sources,  at an optimal network SNR of around 15, 50\% \ of inspiral sources can be located within 23 sq-degs ({\it best case}) at the 95\% confidence level. For the burst source, without any knowledge of the waveform, at SNR of 10, 50\% of the sources can be localized within 50 sq-degs ({\it worst-case}),  but it can be reduced to 8 sq-degs if we have predicted waveforms available (e.g., from Ref. \cite{mario, ott}).  Results for the initial or advanced detectors are similar.  Our results imply that, for prompt follow-up electromagnetic observations directly using triggers from current GW network, wide-field telescopes are desirable.

\acknowledgments  

We are grateful to Emanuele Berti, Cole Miller, Antony Searle, Stephen Fairhurst, Reinhard Prix, Ray Frey,  David Blair, Sterl Phinney, Andrzej Krolak for useful comments of this work. This work is in part supported by NSF grant PHY-0653653, the David and Barbara Groce start-up fund at Caltech, the Alexander von Humboldt Foundation's Sofja Kovalevskaja Programme, and by the Australian Research Council Discovery Grants and Future Fellowship program.


\begin{thebibliography}{}
\bibitem[Kobayashi \& M{\'e}sz{\'a}ros(2003)]{grb_gw} 
Kobayashi, S., \& M{\'e}sz{\'a}ros, P., \apj, {\bf 589}, 861 (2003)


\bibitem[ligo (2009)]{ligo}http://www.ligo.caltech.edu
\bibitem[virgo(2009)]{virgo}http://www.ligo.caltech.edu
\bibitem[geo (2009)]{geo} http://geo600.uni-hannover.de 

\bibitem[Kuroda {et~al.}(1999) ]{lcgt} K.Kuroda et al, Int.J.Mod.Phys.D 8, 557 (1999)

\bibitem[Blair et al.(2008)]{aigo} Blair, D.~G., et al.\ 
2008, Journal of Physics Conference Series, 122, 012001 

  \bibitem[Abbott {et~al.}(2005) ]{ligo_grb} Abbott, B., et al., \prd, {\bf 72}, 042002 (2005) 

\bibitem[Thorne(1987)]{thorne87} Thorne, K. 1987, in ``300 years of Gravitation'', eds Hawking, S.~W.~\& 
Israel, W., Cambridge and New York, Cambridge University Press, 697








\bibitem[G{\"u}rsel \& Tinto(1989)]{tinto89} G{\"u}rsel, Y., \& 
Tinto, M., \prd,{\bf  40}, 3884 (1989)


\bibitem[Wen \& Schutz(2005)]{wen05a} Wen, L., \& Schutz, 
B.~F., CQG, S{\bf 22}, 1321 (2005)

\bibitem[Cavalier et al.(2006)]{cavalier06} Cavalier, F., et al.\ 
2006, \prd, 74, 082004 


\bibitem[Beauville et al.(2006)]{beauville06} Beauville, F., et 
al.\ 2006, Journal of Physics Conference Series, 32, 212 


\bibitem[Acernese et al.(2007)]{acernese07} Acernese, F., et al.\ 
2007, Classical and Quantum Gravity, 24, 617 


\bibitem[Wen(2008)]{wen07a} Wen, L.\ 2008, International 
Journal of Modern Physics D, 17, 1095 

\bibitem[Wen et al.(2008)]{wen_fan} Wen, L., Fan, X., 
\& Chen, Y.\ 2008, Journal of Physics Conference Series, 122, 012038 

\bibitem[Markowitz et al.(2008)]{markowitz08} Markowitz, J., 
Zanolin, M., Cadonati, L., \& Katsavounidis, E.\ 2008, \prd, 78, 122003 


\bibitem[Searle et al.(2008)]{searle08} Searle, A.~C., Sutton, 
P.~J., Tinto, M., 
\& Woan, G.\ 2008, Classical and Quantum Gravity, 25, 114038 


\bibitem[Searle et al.(2009)]{searle09} Searle, A.~C., Sutton, 
P.~J., \& Tinto, M.\ 2009, Classical and Quantum Gravity, 26, 155017 


\bibitem [Fairhurst(2009)]{steve09} Fairhurst, S.\ 2009, New 
Journal of Physics, 11, 123006 




\bibitem[Jaranowski \& Krolak(1994)]{krolak94} Jaranowski, P., 
\& Krolak, A.,\prd, {\bf 49}, 1723 (1994)


\bibitem[Cutler(1998)]{cutler98} Cutler, C., \prd, {\bf 57}, 7089 (1998)  

\bibitem[Pai et al.(2001)]{Pai01} Pai, A., Dhurandhar, S., \& 
Bose, S., \prd, {\bf 64}, 042004 (2001) 

\bibitem[Barack \& Cutler(2004)]{leor04} Barack, L., \& 
Cutler, C., \prd, {\bf 69}, 082005 (2004)

\bibitem[Jaranowski \& Kr{\'o}lak(2005)]{krolak05} Jaranowski, 
P., \& Kr{\'o}lak, A., LRR, {\bf 8}, 3 (2005)


\bibitem[Sylvestre(2003)]{sylvestre03} Sylvestre, J., \apj, 
{\bf 591}, 1152 (2003)



\bibitem[Schutz(1989)]{schutz89} B. F. Schutz, in {\it Gravitational Radiation}, edited by D. Blair, Combrideg University Press, Cambridge, England, 1989


\bibitem[Jaranowski et al.(1998)]{krolak98} Jaranowski, P.,  Kr{\'o}lak, A., \& Schutz, B.~F., \prd, {\bf 58}, 063001 (1998)


\bibitem[Finn(1992)]{finn92} Finn, L.~S., \prd, {\bf 46}, 5236 (1992)


\bibitem[Cramer(1946)]{cramer46}  Cramer, H. 1946, Mathematical Methods of Statistics, Princeton University Press, Princeton, N.J. 


\bibitem[Vallisneri(2008)]{michele08} Vallisneri, M.\ 2008, \prd, 
77, 042001 

\bibitem[Cokelaer(2008)]{tom08} Cokelaer, T.\ 2008, Classical 
and Quantum Gravity, 25, 184007 


\bibitem[Chatterji et al.(2006)]{chatterji06} Chatterji, S., 
Lazzarini, A., Stein, L., Sutton, P.~J., Searle, A., 
\& Tinto, M.\ 2006, \prd, 74, 082005 

\bibitem[Rakhmanov(2006)]{malik06} Rakhmanov, M.\ 2006, 
Classical and Quantum Gravity, 23, 673 



\bibitem[Prix(2007)]{prix07} Prix, R., \prd, {\bf 75}, 023004 (2007) 


\bibitem[Wen et al (2007)]{blair07} Wen, L., Howell, E., Coward, D. and Blair, D.   2007, the XLIInd Rencontres de Moriond, ``2007 Gravitational Waves and Experimental Gravity'', pp 123-130, March 11-16, 2007 - La Thvile, Val d'Aoste, Italie.

\bibitem[Sathyaprakash \& Schutz B. F. (2009)]{schutz09} Sathyaprakash, B. S. \& Schutz, B. F. 2009, Living Review in Relativity 12, 2

\bibitem[Baiotti et al. (2007)]{mario} Luca Baiotti, Roberto De Pietri, Gian Mario Manca and Luciano Rezzolla, 2007 Phys. Rev. D 75, 044023

\bibitem[Ott (2009)]{ott} Ott, C. D 2009, Class. Quan. Grav., 26, 063001


\bibitem[Shoemaker et al.(1996)]{sf_aligo} https://dcc.ligo.org/cgi-bin/DocDB/ShowDocument?docid=2974 

\bibitem[Lazzarini et al.(1996)]{sf_iligo}Lazzarini, A. et al. 1996, LIGO Science Requirement Documents, LIGO~E950018-02 and http://www.ligo.caltech.edu/$~$jzweizig/distribution/LSC\_Data/

\bibitem[Virgo et al. (2009) ] {sf_virgo}http://wwwcascina.virgo.infn.it/senscurve/



\bibitem[Jaranowski et al.(1996)]{pe} Jaranowski, P., 
Kokkotas, K.~D., Kr{\'o}lak, A., 
\& Tsegas, G.\ 1996, CQG, 13, 1279 

\bibitem[LAL (2005)]{lal} LAL Software Documentation, p427




\bibitem[Phinney et al(2009)]{phinney09} Kaswali, M., Phinney, S., Wen, L. and Kulkarni, S. ApJ, to be submitted

\bibitem[Cutler \& Flanagan (1994)]{cutler94} Cutler C. \& Flanagan. E. E. 1994, Phys. Rev. D, 49, 2658 








\end{thebibliography}
\end{document}